\documentclass[reprint,pre,nofootinbib,superscriptaddress]{revtex4-2}

\usepackage[utf8]{inputenc}
\usepackage{amssymb,amsfonts,amsmath,bm}
\usepackage{graphics,graphicx}
\usepackage{hyperref}

\usepackage{braket}
\usepackage{mathtools}
\usepackage{float}
\usepackage{enumitem}
\usepackage{xcolor} 

\DeclarePairedDelimiter\ceil{\lceil}{\rceil} 
\DeclarePairedDelimiter\floor{\lfloor}{\rfloor} 
\newcommand{\e}{{\rm e}} 

\hypersetup{
    colorlinks = true,
    citecolor= black,
    linkcolor=black,
    urlcolor=cyan,
}

\begin{document}

	\title{One-dimensional colloidal model with dielectric inhomogeneity}
	
	\author{Lucas Varela}
	\affiliation{Universit\'e Paris-Saclay, CNRS, LPTMS, 91405, Orsay, France.}	
	\affiliation{Universidad de los Andes, Bogot\'a, Colombia}

	\author{Gabriel T\'ellez}
		\affiliation{Universidad de los Andes, Bogot\'a, Colombia}

\author{Emmanuel Trizac}
\affiliation{Universit\'e Paris-Saclay, CNRS, LPTMS, 91405, Orsay, France.}
		
	\begin{abstract}
  	  We consider a one-dimensional model allowing analytical derivation of the effective interactions between two charged colloids.
       We evaluate exactly the partition function for an electroneutral salt-free suspension  with dielectric jumps at the colloids' position. We derive a contact relation with the pressure that shows there is like-charge attraction, whether or not the counterions are confined between the colloids. In contrast to the homogeneous dielectric case, there is the possibility for the colloids to attract despite the number of counter-ions ($N$) being even. The results are shown to recover the mean-field prediction in the limit $N\to \infty$.
	\end{abstract}

	\maketitle		
	\makeatletter
	\let\toc@pre\relax
	\let\toc@post\relax
	\makeatother 

\section{Introduction}

Electrostatic interactions are key to a wealth of phenomena in soft condensed matter: like-charge attraction,
overcharging/charge inversion, self-assembly, electrophoresis, etc \cite{Holm2001,andelman2006,Levin2002,Naji2005,Boroudjerdi2005,Ioannidou2016}. Nonetheless, understanding 
many-body correlated interactions from a fundamental point of view is usually shielded by mathematical complexities that can only be bypassed with physical insight. Take for example one of the simplest possible settings: two similar charged plates interacting in the presence of neutralizing counter-ions. For high counter-ion valency and/or large colloidal charge, these plates can attract each other providing an example of like-charge attraction.
This phenomenon 
challenges our intuition of electrostatics; it has been reported as early as 1836 in a different setting, for bare like-charged metallic disks at short distances \cite{Harris}. The two previous examples exhibit like-charge attraction through different mechanisms; bare asymmetric conductors have dominant local attractive 
interactions
due to charge redistribution over the surface \cite{Lekner,Alexandre19}, while the plates attract under the mediation of correlated counterions, in a strongly coupled regime
\cite{netzSCtoPB,SaTr11PRL,SaTT18}. This non mean-field effect \cite{Neu1999}
has been confirmed experimentally \cite{Kekicheff1993,crockerGrier,keplerFraden} and computationally  \cite{AllahyarovDamicoLowen,GRONBECHJENSEN,ningGirvinRajaraman,Guldbrand1984,Moreira2002MC}.  While the two phenomena alluded to are ruled
by different mechanisms, both feature short-range attraction, at variance
with the one-dimensional results to be presented in our study, showing 
attraction at long distances.

\begin{figure}[hbt]
	\centering
	\includegraphics[width=0.48\textwidth]{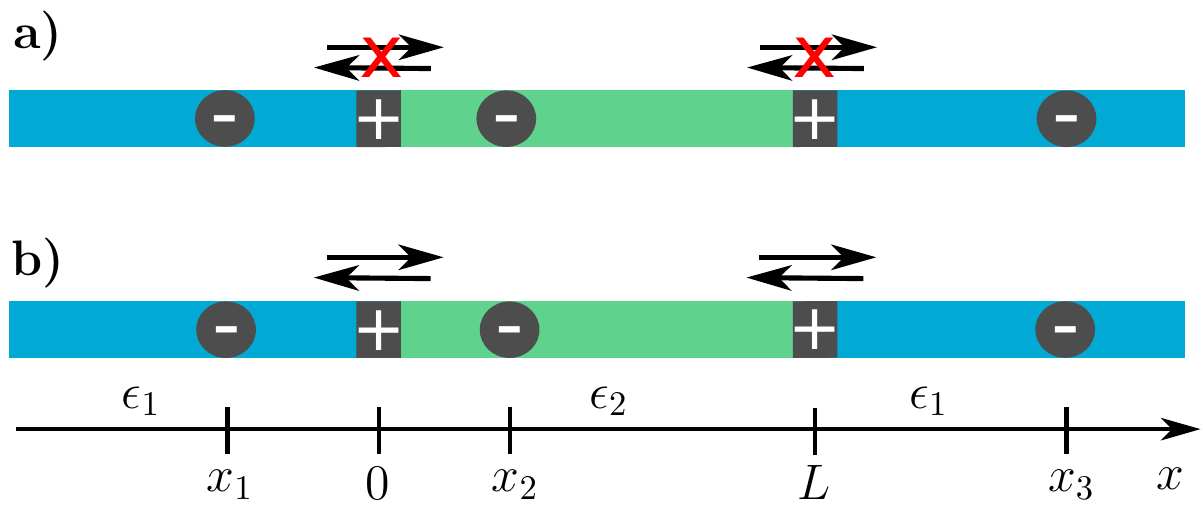}
	\caption{
		 Sketch of the colloidal suspension with 3 counter-ions ($N=3$, shown by the filled disks). The dark  rectangles at $0$ and $L$ represent the colloidal charges. 
		 The dielectric constant is $\epsilon_2$ for $0<x<L$ and $\epsilon_1$ elsewhere. The impermeable case (panel a) forbids particles to cross the spatial regions delimited by the colloids' positions. For example, in sketch a) the {positions $x_1,x_2$ and $x_3$ of the left, center and right counterion are  restricted to the intervals given by $x_1<0,\, 0<x_2<L$, and $x_3>L$ respectively.} Instead, the permeability condition (panel b) refers to counter-ions with no restriction on {the positions $x_1,x_2$ and $x_3$}. In the canonical situation, the distance $L$ between the two colloids is fixed while it does fluctuate under isobaric 
		 conditions.
	}
	\label{sketch}	
\end{figure}

The majority of earlier studies concentrated on systems with a global homogeneous dielectric medium. However, while a relevant element in the description of colloidal systems lies in the dielectric discontinuity between the inside of the colloidal particles and the solvent medium, exact results accounting for this effect are scarce \cite{Jho2008,Kjellander1984,Kanduc2007,SaTr2012EPL,SaTr2012CPP,Samaj2016}. The present work reports exact results in a one-dimensional setting for the pressure and counter-ion density profile in the presence of dielectric jumps. Inspiration comes from previous studies which used lower dimensional systems for both colloids \cite{Dean1998,Dean2009,demery2012,tt2015,frydel2019} 
and electrolytes \cite{lenard,lenard2,prager,baxter} to compute the properties for a homogeneous dielectric medium. We consider here two symmetric colloids and $N$ neutralizing counter-ions, a so-called salt-free system (no counter-ions). Unlike in our previous work \cite{vtt2016}, the underlying medium is not a homogeneous dielectric, but features a dielectric discontinuity. The considered piece-wise linear dielectric medium changes the interaction potential. At each colloid's boundary  there is a dielectric discontinuity,  as seen in Fig.  \ref{sketch}. Two models are considered: the
colloids' boundary are either impermeable to the counter-ions, or not. When they are permeable,
the counter-ions not only populate the interstitial region between the colloids 
($0\leq x \leq L$ in Fig. \ref{sketch}),
but also the colloid's interior ($x\leq 0$
and $x\geq L$), see e.g. \cite{CMTR09,CCRT11,BaTr12}
for an analysis of this penetrable model.

It is worth stressing here that one-dimensional models may shed {light} on more realistic three dimensional situations. Indeed, the equation of state for two like-charged plates with neutralizing counterions has been shown to coincide, under conditions of high enough coupling and small separations, to its one dimensional counterpart \cite{Moreira2002MC,SaTr11PRL,Varenna}. This stems from the fact that when the inter-plate separation is smaller than the typical distance between counterions, counterion interactions become immaterial and the 1D model with $N=1$ counterion subsumes the 3D phenomenology.
We will start by investigating this single counterion case. 

The paper is organized as follows. 
The ``pressure'', which in 1D is nothing but the force, is obtained in section \ref{sec:1counterion} for both canonical and isobaric ensembles to examine the possible occurrence of like-charge attraction. The  density profile is then calculated and a contact condition is established for both impermeable and permeable colloids. In the former case, the effects of electrostatic images cancel out and we get the same result as for the homogeneous dielectric situation,  while for the latter, the dielectric jump 
does modulate the counter-ion density and consequently the pressure. In section \ref{sec:N}, the one particle results are generalized for an arbitrary number $N$ of interacting counter-ions. Finally, we compare in section \ref{sec:mf} our exact treatment to the mean-field prediction, which proves to be a fair approximation even for small $N$.


\section{One Counter-ion}
\label{sec:1counterion}

In this section, we present the results for a single counter-ion  ($N=1$) of charge $e$. Such a limiting case is distant from experimental reality, where the colloidal charge largely exceeds the ionic one; it nevertheless offers a useful starting point.
Two colloids, each with charge $-e/2$, are located at $x = 0$ and $x = L$. The dielectric medium is piece-wise constant with $\epsilon_2$ for $x\in [0,L]$ and $\epsilon_1$ everywhere else (Fig.  \ref{sketch}). We characterize the dielectric jump  by the parameter $\Delta =(\epsilon_2-\epsilon_1)/(\epsilon_2+\epsilon_1)$. This quantity is bounded, with $-1\leq \Delta \leq 1$.  The two dielectric jumps create an infinite set of image charges, leading to an electric potential defined in equations \eqref{potl}-\eqref{potm}. Note that the self energy terms $q q'\Delta L/(1-\Delta )^2 \epsilon_2$ induced by the image charges will cancel out in the total potential energy due to electroneutrality, unlike the cases with a one discontinuity in 1D or an arbitrary number in 3D.
This section first addresses the impermeable case, and then connects it to its permeable counterpart. In the former, the counter-ion is restricted to be in one of the three spatial regions delimited by the points where the dielectric medium is discontinuous. 
On the other hand, in the permeable situation, the counterion  position is unrestricted. In both cases, we obtain a contact-theorem-like relation, establishing a connection between the inter colloidal force, and some contact density.
    
\subsection{Impermeable Colloids\label{subsec:1ci}}
    
With an impenetrable wall at $0$ and $L$, there are three possibilities to place the counter-ion: $x<0, 0<x<L$ or $x>L$. Throughout this paper, we will call these the left, middle and right regions respectively. We can then label each configuration by specifying the total number of counter-ions $N$ and the sub-indexes $N_{\ell}$ and $N_r$ specify how many counter-ions are in the left and right regions respectively. With this notation, the potential energy $U_1$ for a colloidal suspension with  one counterion is: 
\begin{equation}
U_1 = \frac{e^2}{\epsilon_2} \left[  \frac{L}{4} - L N_r \left( \frac{1+\Delta}{1-\Delta}\right)   + x(N_r-N_{\ell})\left(\frac{1+\Delta}{1-\Delta}\right)\right].
\label{U1}
\end{equation}
The previous expression is obtained by adding the potential interactions $V(x,x')$ among particles and the self interaction terms $V(x,x)/2$, where $V$ is defined in Appendix \ref{appendix:potential}.
The counter-ion density is proportional to the Boltzmann factor of $U_1$ and the partition functions $z_{N_{\ell},N_r}(N,\widetilde{L},\Delta)$ for each possible impermeable systems follow as:
\begin{subequations}
	\begin{align}
	z_{1,0}(1,\widetilde{L},\Delta) &=\left(\frac{1-\Delta}{1+\Delta}\right) \e^{-\widetilde{L}/4} & \text{left region} \\
	z_{0,0}(1,\widetilde{L},\Delta) &= \widetilde{L}\; \e^{-\widetilde{L}/4} &  \text{middle region}\\
	z_{0,1}(1,\widetilde{L},\Delta) &= \left(\frac{1-\Delta}{1+\Delta}\right) \e^{-\widetilde{L}/4} &  \text{right region}
	\end{align}
	\label{z1}
\end{subequations}
where  $\widetilde{x} = x\beta e^2/\epsilon_2$ ($\beta = 1/k_BT$) is a dimensionless length. These partition functions are proportional to their homogeneous expressions with $\Delta =0$:
\begin{equation}
z_{N_{\ell},N_r}(1,\widetilde{L},\Delta) = \left(\frac{\epsilon_1}{\epsilon_2}\right)^{N_{\ell} + N_r}z_{N_{\ell},N_r}(1,\widetilde{L},0),
\label{prop1}
\end{equation}
where $\epsilon_1/\epsilon_2 = (1-\Delta)/(1+\Delta)$. Equation \eqref{prop1} yields the same pressure as for a system with $\Delta=0$, which was shown in \cite{vtt2016} to follow the form of the contact condition \cite{henderson1,henderson2},
where a pressure is written as the sum of a contact density term, minus a term involving the square of some charge
\begin{subequations}
	\begin{align}
	\widetilde{n}(0^+) &= \widetilde{P}_c+\left(1/2-N_{\ell}\right)^2 
	\label{CT_1c_1} \\
	\widetilde{n}(0^-) &= \left(\frac{\epsilon_2}{\epsilon_1}\right)N_{\ell}^2 .
	\label{CT_1c_2}
	\end{align} 	
\end{subequations}
Here, $\widetilde{P}_c = \epsilon_2P_c/e^2$ and  $\widetilde{n} = n \beta e^2/\epsilon_2$ are the rescaled canonical pressure and particle number density respectively and the discontinuity of the ionic density at $x=0$ is a consequence of the impermeability of the three compartments. Recovering the contact relation is not a surprise, since it is an exact result for impermeable 
charged bodies.
Equation \eqref{CT_1c_2} is the contact theorem in the left region ($\widetilde{x} \leq 0)$: this region does not contribute to the force exerted on the left-most charge. 
Besides, the factor $\epsilon_2/\epsilon_1$  stems from the dimensionless number density $\widetilde{n} = \epsilon_2 n/\beta e^2$. The equivalent expressions for $\widetilde{n}(\widetilde{L}^{\pm})$ follow from the replacement $N_{\ell} \to N_r$.
The particular case where the counter-ion is between the colloids has been studied before and it will be shown to be equivalent to the permeable  situation with $\Delta = 1$, where the ion cannot escape the central segment $0<x<L$, due to the strong image charge repulsion. It can be shown from equation \eqref{CT_1c_1} and the expression for $\widetilde{n}(0^+)$ that follows from \eqref{U1} and \eqref{z1}, that independently from $L$, there is like-charge attraction. As alluded to in the introduction,  the phenomenology obtained in 1D with $N=1$ counterion 
is relevant for the prototypical 3D system of strongly charged plates neutralized by counterions: the small distance equation of state takes the very same form in both cases, as discussed in \cite{Varenna}.
In the impermeable case, we conclude that the dielectric jump is invisible to the pressure, from  a cancellation of the forces exerted by the (in 1D infinite range) image charges. Each dielectric only affects the distribution of the particles that occupy it and the sole contribution to the colloid's pressure is exclusively through the total charge of the left and right regions. 
    
\subsection{Permeable Colloids\label{subsec:1cp}}
    
We now allow the counter-ion of charge $e$ to {lie} anywhere in the line, without positional restriction. Unlike with 3D Coulomb potential, there is no divergent term when the counter-ion overlaps with the colloidal (point) charge, due to the linear nature of the 1D Coulomb potential. This situation is equivalent to that in three dimensions, when a point ion approaches a uniformly charged plane.
We move on to describe in detail the equation of state, the contact condition and the counter-ion position fluctuations. 

\subsubsection{Equation of state}

\begin{figure}[ht]
	\centering
	\includegraphics[width=0.48\textwidth]{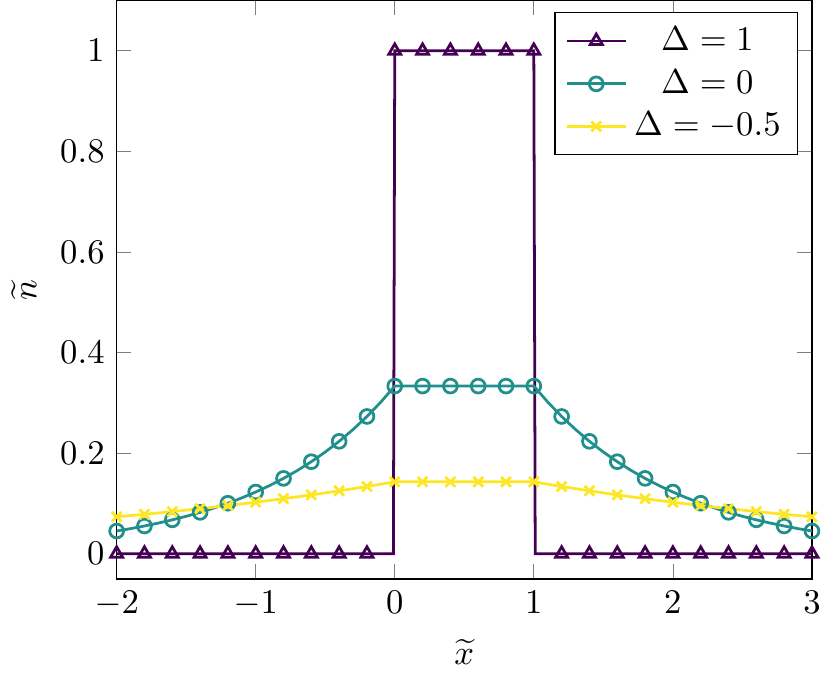}%
	\caption{
		Number density for a one ($N=1$) counter-ion system with permeable colloids at a distance $\widetilde{L}=1$. The probability of finding the particle between the colloids is monotonically increasing with $\Delta$. By virtue of the contact theorem, the repulsive contribution to the pressure increases, as so does the pressure itself, as seen in Fig. \ref{fig:P_1}. The limit $\Delta = 1$ is equivalent to the model studied in \cite{Dean2009,tt2015} where two {\em impermeable} colloids interact with counter-ions sandwiched in between.}
	\label{fig:densityN1}
\end{figure}

Getting  physical intuition on the pressure's behavior requires an understanding of how the counter-ion number density $n_1$ is shaped by the dielectric jump. This connection is encoded in the contact theorem, but it can also be identified with the expression obtained through the direct calculation of the pressure, following from the free energy. 
In the present one particle problem, the ionic density is again given by the Boltzmann factor of the potential in
equation \eqref{U1}
\begin{equation}
\widetilde{n}_1(\widetilde{x},\widetilde{L},\Delta) =\frac{1}{\widetilde{L} +2\left(\frac{1-\Delta}{1+\Delta}\right) } 
\begin{cases}
\e^{\widetilde{x} \left(\frac{1+\Delta}{1-\Delta}\right)} & \widetilde{x}<0\\
1 & 0<\widetilde{x}<\widetilde{L}\\
\e^{(\widetilde{L}-\widetilde{x}) \left(\frac{1+\Delta}{1-\Delta}\right)} & \widetilde{x}>\widetilde{L},
\end{cases}
\label{1c_n}
\end{equation}
where the continuity of the ionic density is enforced. We observe the appearance of a decay length $(1-\Delta) /(1+\Delta) = \epsilon_1/\epsilon_2$ at each side,
quantifying the ``leaking'' of the ion outside the central region. 
Equation \eqref{1c_n} shows that for negative dielectric jumps ($\epsilon_1>\epsilon_2$), attractive images drive the counter-ion to the exterior regions (left and right). In contrast, when $\Delta >0$ the image charges repel the counter-ion thus increasing the middle region density (see Fig. \ref{fig:densityN1}). This leads to an increase of pressure as a function of $\Delta$.

For a system at constant length (canonical situation), the permeable and impermeable cases are related through their partition functions. The former is given by the sum of all the latter situations: $Z_c(1,\widetilde{L},\Delta) =		z_{1,0}(1,\widetilde{L},\Delta)  +	z_{0,0}(1,\widetilde{L},\Delta) + 	z_{0,1}(1,\widetilde{L},\Delta)$. For $N = 1$, a direct computation is tractable and thus it can be checked by direct integration of the Boltzmann factor that $Z_c(1,\widetilde{L},\Delta)$ is the sum of all possible $z_{N_{\ell},N_r}(1,\widetilde{L},\Delta)$:
\begin{equation}
Z_c(1,\widetilde{L},\Delta) = \e^{-\widetilde{L}/4} \left[\widetilde{L}+2\left(\frac{ 1-\Delta }{1+\Delta}\right)\right].
\label{1c_zc}
\end{equation} 
It follows that the canonical pressure $\widetilde{P}_c = d \ln Z_c / dL$ can be written as the sum of an attractive and a repulsive term
\begin{equation}
\widetilde{P_c} = -\frac{1}{4} + \frac{1}{\widetilde{L} +2\left(\frac{1-\Delta}{1+\Delta}\right) } .
\label{1c_Pc}
\end{equation}

\begin{figure}[ht]
	\centering
	\includegraphics[width=0.48\textwidth]{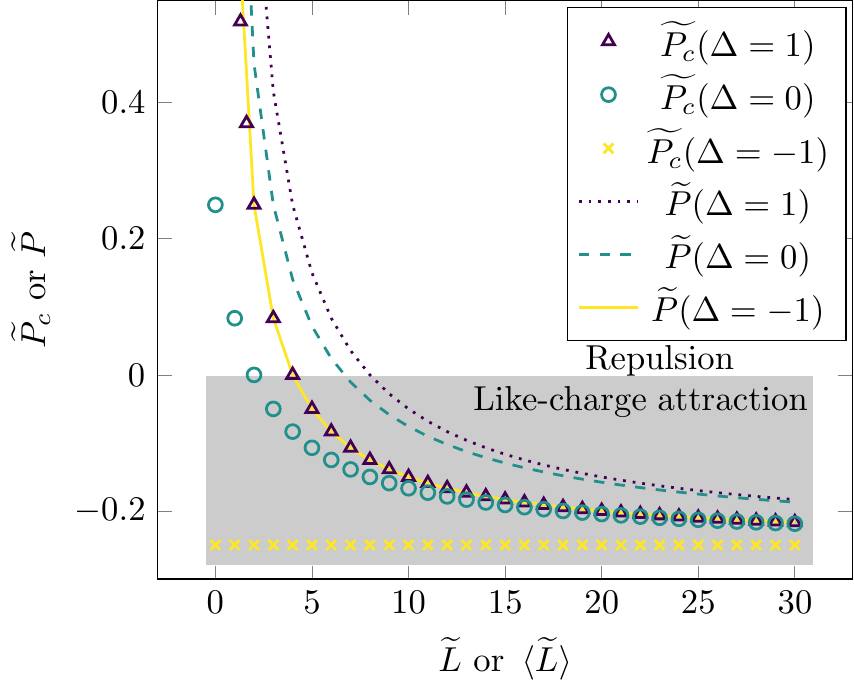}%
	\caption{
		Equation of state for the canonical ($\widetilde{P}_c$) and isobaric ($\widetilde{P}$) ensemble with a permeable colloids and $N=1$ counter-ion. Except for $\Delta = 1$, the effective length $L_{\text{eff}}$ is non vanishing and thus keeps the canonical pressure bounded for any colloid distance, at variance with the average isobaric length $\braket{\widetilde{L}}$ that vanishes as $\widetilde{P} \to \infty$.
		The extreme dielectric
		discontinuity with $\Delta=1$ indeed makes ionic excursions in the outer right or left regions energetically  too costly. There is the possibility of like-charge attraction (state points inside the gray shade) for any given dielectric jump $\Delta$ since both $\widetilde{P}_c$ and $\widetilde{P}$ tend to $-1/4$ at infinite colloid separation.  The reason why $\widetilde{P}_c(\Delta = 1)= \widetilde{P}(\Delta = -1)$ is explained in Appendix \ref{appx:deltaequivalance}.}
	\label{fig:P_1}
\end{figure}

 The attractive term is the force between two opposite charges $\pm e/2$ while the repulsive term is the pressure exerted by a free counter-ion confined in an effective length $L_{\text{eff}} =  \widetilde{L} + 2 \epsilon_1/\epsilon_2$. The counterion density in the middle region is indeed uniform, for the reason that the
electric field acting there does cancel by symmetry. 
The effective length is the sum of the colloids' distance and the decay length alluded to
after Eq. \eqref{1c_n}. Since there is an $\widetilde{L}$ independent term in the effective length, the pressure remains finite even when the colloids collapse onto each other ($\widetilde{L}\to 0$). 
The exception is for $\Delta=1$, where the effective length vanishes since the model becomes {effectively} impermeable: the counterion cannot escape the middle region, which leads to a diverging entropic cost for $L\to 0$ in the canonical fixed-$L$ ensemble, and thus a diverging pressure. {The confinement of the counterion at $\Delta =1$  can be understood in terms of image charge interactions: for $\Delta>0$  the dielectric jump determines the magnitude of the repulsive force exerted onto the counterion by the image charges, which at $\Delta=1$ is maximal and strong enough to prevent the counterion from leaving the middle region. In this sense, the system becomes effectively impermeable due to the confinement effect of the image charges, and not because the colloids would preclude the ions to go through.}
Fig. \ref{fig:P_1} shows that a region of  like-charge attraction always exists regardless of the dielectric jump. It is given by $\widetilde{L}> (2+6\Delta)/(1+\Delta)$. Remembering the definition
of our rescaled lengths, $\widetilde L = L \beta e^2/\epsilon_2$, this criterion is expected: it states that for a fixed length $L$, attraction is triggered by decreasing the temperature $T\propto \beta^{-1}$: like-charge attraction indeed is a strong coupling phenomenon, here a low-$T$ feature.

The contact theorem is derived for an impermeable system but an equivalent relation can be found for permeable situations. By replacing the electric field by its statistical average, we get the contact condition:
\begin{equation}
\widetilde{P_c} = \widetilde{n}(0) - \left<\left(\frac{1}{2}-N_{\ell}\right)^2 \right>,
\label{1c_CT}
\end{equation}
where the term in the average is the square of the total charge in the left region $\widetilde{x}\leq0$, including thus the charge $-e/2$ at $x=0$.
The contact theorem yields equation \eqref{1c_Pc} when the quantities $\widetilde{n}(0)$, $\braket{N_{\ell}}$ and $\braket{N_{\ell}^2}$ are replaced by their explicit expressions, which will be computed in the next section. 
Therefore, we can view the (positive) repulsive term in equation \eqref{1c_CT} as given by the contact number density which is monotonically increasing with $\Delta$. As previously stated, the attractive term is constant and given by the force between the colloidal particles. This force follows from the term $-(1/2 - N_{\ell})^2$, which yields $-1/4$ due to the special feature $N_{\ell} = N_{\ell}^2$ of the single counter-ion case ($N_\ell$ is indeed either 0 or 1). Note that Eq. \eqref{1c_CT} when $\Delta = 1$ coincides with Eq. \eqref{CT_1c_1} because these respective permeable and impermeable cases have the same density profile: $\Delta=1$ precludes counterions excursions outside the central region.\\

We now turn to the equation of state for the isobaric ensemble. 
{In this ensemble the number of particles, temperature and pressure are fixed. Unlike in the canonical case, the system's length is not fixed; it is allowed to fluctuate in order to preserve a constant pressure by exchanging length and work with the barostat (i.e. the piston).
Then, instead of using the system's length for the equation of state we compute its average, which} is given by $ \braket{\widetilde{L}} = -\partial \ln Z_P/ \partial \widetilde{P}$, where $Z_P$ is the Laplace transform of the canonical partition function:
\begin{equation}
\braket{\widetilde{L}}= \frac{2}{\widetilde{P}+1/4}-\frac{4 (1-\Delta)}{4(1-\Delta) \widetilde{P}+3+\Delta}.
\label{isobariclength1}
\end{equation}
The inversion of $\widetilde{P}$ as a function of $\braket{\widetilde{L}}$ shows that the asymptotic value of the pressure at infinite colloid distance is $-1/4$, which is the same as the canonical ensemble limit. 
        
\subsubsection{Fluctuations}

It was previously stated that the fluctuations are irrelevant to understand the pressure for $N=1$ because all moments of the number of left (and right) counter-ions are the same $\braket{N_{\ell}^m} = \braket{N_{\ell}}$ for all orders $m$.  
However, for $N>1$, they will play a key role and for the sake of completeness, we discuss fluctuations already for $N = 1$. The moments $\braket{N_{\ell}^m}$ can be computed using the probability of each of the impermeable configurations: $p_{N_{\ell},N_r} = z_{N_{\ell},N_r}/Z$. The average is then defined as $\braket{(\cdot)} = \sum_{N_{\ell},N_r} (\cdot) p_{N_{\ell},N_r}$. In the present case, $\braket{N_{\ell}}$ only features a contribution from $z_{1,0}/Z$:
\begin{equation}
	\braket{N_{\ell}} = \braket{N_{\ell}^2}	= \frac{\frac{1-\Delta}{1+\Delta}}{\widetilde{L} +2\left(\frac{1-\Delta}{1+\Delta}\right) },
	\label{avg_Nl_N-1}
\end{equation}
so that			
\begin{equation}
\frac{\left<N_{\ell}^2\right>}{\left<N_{\ell}\right>^2} = 2+\left(\frac{1+\Delta}{1-\Delta}\right)\widetilde{L} .
	\label{1i-fluc}
\end{equation}
The fluctuations of $N_\ell$ are monotonically increasing in $\widetilde{L}$, with a range given by $2\leq \left<N_{\ell}^2\right> / \left<N_{\ell}\right>^2 < \infty$. For greater number of counter-ions, the behavior is also monotonically increasing, but unlike for $N=1$,  $\left<N_{\ell}^2\right> / \left<N_{\ell}\right>^2$ is bounded from above. 

We now turn our attention to the compressibility and its relation to the variance of the number of particles, $\sigma_N^2$. In a grand canonical situation, $\sigma^2_N$ would be related to the compressibility $\chi_T$ through 
 \begin{equation}
  k_B T\chi_T = \frac{L\,\sigma_N^2}{\braket{N}^2} \ .
 \label{eq:fluctresp}
 \end{equation}
 We should not expect this fluctuation-response connection to hold in our canonical or isobaric cases; it is nevertheless instructive to study the quantitative violation 
 of this relation. First, we compute the variance of the number of inside counter-ions $N_{\text{in}} = N-N_{\ell} - N_{r}$:
\begin{equation}
	\sigma_{N_{\text{in}}}^2  = \left(\frac{ \widetilde{L} }{\widetilde{L}+2 \left(\frac{1-\Delta}{1+\Delta}\right)} \right) \left(\frac{2\left(\frac{1-\Delta}{1+\Delta}\right)}{\widetilde{L}+2 \left(\frac{1-\Delta}{1+\Delta}\right)}\right),
\end{equation}
where we identify in the right hand side of the equation two factors: $\braket{N_{\text{in}}}$ (left factor) and $\braket{N_{\text{out}}}$ (where $N_{\text{out}} = N_{\ell} + N_{r}$) in the right. We then get $\sigma_{N_{\text{in}}}^2 =  \braket{N_{\text{in}}}\braket{N_{\text{out}}}$. Using the expression for $\braket{N_{\text{in}}}$  follows that $\sigma_{N_{\text{in}}}^2/\braket{N_{\text{in}}} \rho_{\text{in}}$ is the total exterior effective length:
\begin{equation}
\widetilde{L}\frac{\sigma_{N_{\text{in}}}^2}{\braket{N_{\text{in}}}^2} = 2 \left(\frac{1-\Delta}{1+\Delta}\right),
\end{equation}
where we have used $\rho_{\text{in}} =  \braket{N_{\text{in}}} /L$. To see how this compares to the direct calculation of the compressibility, we proceed to compute  $\widetilde{\chi}_c^{-1} = -\widetilde{L}\partial \widetilde{P_c} /\partial \widetilde{L}$: 
\begin{equation}
	\widetilde{\chi}_c =	\frac{4}{\widetilde{L}} \left(\frac{1-\Delta}{1+\Delta} \right)^2+ 4 \left(\frac{1-\Delta}{1+\Delta}\right) + \widetilde{L} .
	\label{chi_c}
\end{equation}
 We identify in the previous equation the compressibility of the impermeable configuration with $N_{\ell} = N_r = 0$, $\widetilde{L}$, which is dominant for large lengths. Quite expectedly, this very term is recovered with $\Delta=1$.
 The computations for the isobaric ensemble are analog:
\begin{equation}
\braket{\widetilde{L}}\frac{\sigma^2_{P_{N_{\text{in}}}}}{\braket{N_{\text{in}}}_P^2}=  	\braket{\widetilde{L}}\left(\frac{1-\Delta}{1+\Delta}\right)\left( 2 \widetilde{P} + \frac{1}{2} \right),
\end{equation}
where $\braket{\widetilde{L}}$ is given by Eq. \eqref{isobariclength1}. The isobaric compressibility $\widetilde{\chi}_P = -\braket{\widetilde{L}}^{-1} \partial \braket{\widetilde{L}}/\partial \widetilde{P}$ is:
\begin{equation}
	\widetilde{\chi}_P =\frac{1}{ \widetilde{P}+\frac{1}{4}}+\frac{4 (1-\Delta)}{4\widetilde{P}(1- \Delta ) +3+\Delta}-\frac{4 (1-\Delta)}{4\widetilde{P}(1- \Delta ) +5+\Delta}.
\end{equation}

\begin{figure}[ht]
	\centering
	\includegraphics[width=0.46\textwidth]{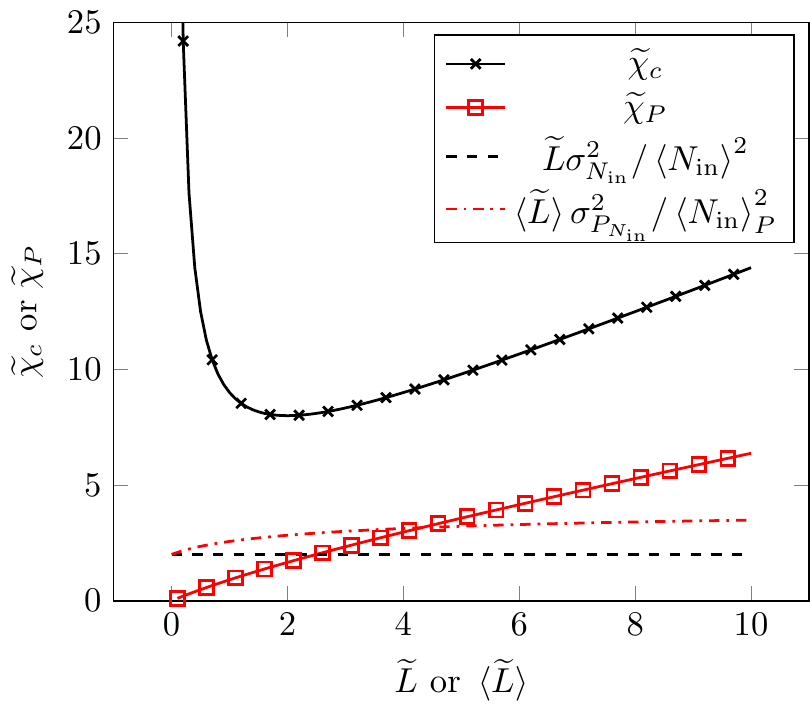}%
	\caption{
			Canonical and isobaric compressibilities as a function of $\widetilde{L}$ and $\braket{\widetilde{L}}$ respectively, at $\Delta = 0$. The fluctuation-response connection given by Eq.~\eqref{eq:fluctresp} (broken lines, one for canonical, one for isobaric)  is clearly violated in both ensembles. The behavior at zero length is radically  different: $\widetilde{\chi}_c$ diverges while $\widetilde{\chi}_P$ vanishes. 
			Both ensembles predict unbounded growth with differing rates as $\widetilde{L} \to \infty$.}
	\label{fig:chiN1}
\end{figure}

Just as in the canonical ensemble, the term  for the impermeable colloid with the counterion in the interstitial region appears  explicitly, $1/(\widetilde{P}+1/4)$, and it  dominates for $\widetilde{P} \to -1/4$.
In order to compare with the canonical ensemble results, we express the isobaric pressure as a function of $\braket{\widetilde{L}}$ and find that $\widetilde{\chi}_P$ vanishes as $\braket{\widetilde{L}} \to 0$. This is at variance with the canonical expression, Eq. \eqref{chi_c}, which diverges as $1/\widetilde{L}$, see Fig. \ref{fig:chiN1}.
This is understood as follows. 
The isobaric length of the system requires an infinite pressure to vanish, and be fluctuation-less. 
The corresponding susceptibility, measured from the response of the mean length to an extra change of pressure, thus vanishes.
On the other hand, at a fixed $\widetilde{L}$ close to 0,   
$\partial \widetilde{L}/\partial \widetilde{P}_c$ approaches its minimum non-vanishing value due to the decay length term and thus  $\widetilde{L}^{-1}\partial \widetilde{L}/\partial \widetilde{P}_c$ diverges. In other words, 
within the canonical description at short separations, the mean number of counterions in the interior regions is small, and does not resist compression. Hence, the large compressibility, signalled by the divergence of $\widetilde \chi_c$.
Besides, in the infinite length limit, both compressibilities show linear behavior: $\lim_{\braket{\widetilde{L}}\to \infty} \widetilde{\chi}_P/\braket{\widetilde{L}}  = \left( 2+\Delta \right)/4$  and $\lim_{\widetilde{L}\to\infty}\widetilde{\chi}_c/\widetilde{L} = 1$. This illustrates how ensembles drastically differ, when the thermodynamic limit is not being considered (note that 
$\widetilde L \to \infty$ does not correspond to the thermodynamic limit since $N$ is here fixed to one).
 
\section{N counter-ions\label{sec:N}}
Let us now consider the case of $N$ counter-ions of charge $e$ in the same dielectric setting as considered so far, and colloids with charge $-eN/2$ each such that electroneutrality holds. We will compute the equation of state for any $N$ and show, both from  statistical and mechanical arguments, that a contact condition exists regardless of the permeability of the colloids. The electrostatic potential energy $U_N$ for a colloidal suspension with $N$ counterions reads:
\begin{align}
&U_N = \frac{e^2L}{\epsilon_2} \left[\frac{N^2}{4} - N N_r - \frac{2\Delta N_r^2}{1-\Delta}  \right] + e^2 \left[\sum_{i=1}^{N_{\ell}} (1-2i) \frac{x_i}{\epsilon_1} \right. \nonumber\\
&\left.+ \sum_{i=N_{\ell}+1}^{N-N_r}(2i-1-N)\frac{x_i}{\epsilon_2} + \sum_{i=N-N_r + 1}^{N}(2N - 2i + 1)\frac{x_i}{\epsilon_1} \right],
\label{UN}
\end{align}
where the positions are labeled such that $\widetilde{x}_1 < \cdots< x_{\ell} <0 <x_{\ell+1} <\cdots  <x_{N-N_r}< \widetilde{L} < x_{N-N_r+1} < \cdots< \widetilde{x}_N$ and $U_N$ is computed using the same procedure as outlined for $U_1$ (see Eq. \eqref{U1}). The first $N_{\ell}$ and last $N_{r}$ positions are in the left and right regions respectively. From the previous equation the force felt by  particle  $k$ is:
\begin{equation}
- \frac{\partial U_N}{\partial x_k}  = 
\begin{dcases}
	\frac{e^2}{\epsilon_1}(2k-1) & k\leq N_{\ell}\\
	\frac{e^2}{\epsilon_2}(N+1-2k) & N_{\ell}<k \leq N- N_r\\
	\frac{e^2}{\epsilon_1}(2k-2N-1) & k>N-N_r.
\end{dcases}
\end{equation}

Just as in the case $\Delta = 0$, the force felt by a counter-ion in 1D depends only on the difference between the total charge on its right ($e(N-k)$) and that on its left ($e(k-1)$). However, due to the factors $\epsilon_{1,2}$, the forces on the counter-ions are not commensurable in general. 

\subsection{Impermeable Colloids}
\label{sec:Nimpermeable}

In this section, we will derive results for an impermeable colloid with an arbitrary number of counter-ions, which as in the one counter-ion case will provide the building block of the partition function in the permeable case.  
The system consists of $N$ counter-ions with a fixed number of particles in each of the three regions: $N_{\ell}$, $N-N_{\ell}-N_r$ and $N_r$ in the left, middle and right region respectively. With this convention, the partition function $z_{N_{\ell},N_r} (N,\Delta,\widetilde{L})$ is:
\begin{align}
	z_{N_{\ell},N_r} =\int_{ \mathcal{D}} \e^{-\widetilde{U}_N}\prod_{k=1}^{N} d\widetilde{x}_k, 
	\label{z_imperm}
\end{align}
where $\widetilde{U}_N= U_N \beta$ and $\mathcal{D} = \{-\infty< \widetilde{x}_1 < \cdots< x_{\ell} <0 <x_{\ell+1} <\cdots  <x_{N-N_r}< \widetilde{L} < x_{N-N_r+1} < \cdots< \widetilde{x}_N < \infty\}$. Following a similar procedure as in previous works  \cite{lenard,tt2015,vtt2016}, we find that the partition function is proportional to the uniform dielectric expression ($\Delta = 0$):	\begin{equation}
	z_{N_{\ell},N_r}(N,\Delta,\widetilde{L}) = \left(\frac{1-\Delta}{1+\Delta}\right)^{N_{\ell}+N_r} z_{N_{\ell},N_r}(N,0,\widetilde{L}).
\end{equation}
where $z_{N_{\ell},N_r}(N,0,\widetilde{L})$ is found in \cite{vtt2016}. As stated in section \ref{sec:1counterion} and discussed in Appendix \ref{sec:CT}, there is a contact condition for the uniform dielectric case, that transposes here into
\begin{subequations}
	\begin{align}
		\widetilde{n}_{N_{\ell},N_r}(0^+) &= \widetilde{P}_c+\left(N/2-N_{\ell}\right)^2,\label{CT_N_1}\\
		\widetilde{n}_{N_{\ell},N_r}(0^-) &= \left(\frac{\epsilon_2}{\epsilon_1}\right)N_{\ell}^2,
		\label{CT_N_2}
	\end{align} 	
\end{subequations}  
where the density profile is given by
\begin{align}
&\widetilde{n}_{N_{\ell},N_r}(\widetilde{x}<0)=
\frac{N_{\ell}!^2(1+\Delta)}{(1-\Delta)}\nonumber\\
&\times\sum_{k=1}^{N_{\ell}}\sum_{j=k}^{N_{\ell}} \frac{ 2 j(-1)^{2N_{\ell}-k-j}  (j+k-1)! \e^{j^2 \widetilde{x}\left(\frac{1+\Delta}{1-\Delta}\right)}}{(k-1)!^2 (j+N_{\ell})! (j-k)! (N_{\ell}-j)!}
\label{imperm_n1}
\end{align}

\begin{align}
&\widetilde{n}_{N_{\ell},N_r}(0<\widetilde{x}<\widetilde{L})=\frac{\left(\frac{1+\Delta}{1-\Delta}\right)^{N+1}}{z_{N_{\ell},N_r}(N,\Delta ,L)}\sum _{k=N_{\ell}+1}^{N-N_{r}}  k!^2 \nonumber\\
& \times(N+1-k)!^2
z_{N_{\ell},N+1-k}(N,\Delta ,x) z_{k,N_r}(N,\Delta ,L-x)
\label{imperm_n2}
\end{align}
and $\widetilde{n}_{N_{\ell},N_r}(\widetilde{x}>\widetilde{L}) = \widetilde{n}_{N_r,N_{\ell}}(\widetilde{L}-\widetilde{x}) $.

\subsection{Permeable Colloids}
\label{subsec:Ncp}

We generalize next the permeable case introduced in section \ref{subsec:1cp} to an arbitrary number of ions $N$. The canonical partition function can be written again in terms of nested integrals:
\begin{align}
	Z_c(N,\Delta,\widetilde{L}) = \int_{-\infty< \widetilde{x}_1 < \cdots < \widetilde{x}_N < \infty }\; \e^{-\widetilde{U}_N}\prod_{k=1}^{N} d\widetilde{x}_k. 
	\label{zc_N}  
\end{align}
Just as in the one counter-ion case, equation \eqref{zc_N} is the sum over all the possible configurations of $N$ particles arranged in the three regions delimited by the colloid's positions:
\begin{align}
	Z_c(N,\Delta,\widetilde{L}) &= \sum_{N_{\ell} = 0}^{N}\sum_{N_{r} = 0}^{N-N_{\ell}} z_{N_{\ell},N_r}(N,\Delta,\widetilde{L}), \\
	Z_P(N,\Delta,\widetilde{P}) &= \sum_{N_{\ell} = 0}^{N}\sum_{N_{r} = 0}^{N-N_{\ell}} {z_P}_{N_{\ell},N_r}(N,\Delta,\widetilde{P}),
\end{align}
where the lowercase partition functions have $N_{\ell}$ counter-ions at $\widetilde{x}<0$ and $N_r$ at $\widetilde{x}>\widetilde{L}$. For the isobaric partition function, we reach a compact expression:

	\begin{align}
&Z_P = \left(\sum_{n=0}^{\left\lceil \frac{N}{2}\right\rceil -1} \sum_{m=0}^{n}+\sum_{n=\left\lceil \frac{N}{2}\right\rceil}^{N} \sum_{m=0}^{N-n}\right) \frac{2\left(\frac{1-\Delta}{1+\Delta}\right)^{n+m}}{2^{\delta_{mn}}(n!m!)^2}  \nonumber \\\times& \frac{\Gamma (m-\frac{N}{2}-i \sqrt{\widetilde{P}}) \Gamma(m-\frac{N}{2}+i \sqrt{\widetilde{P}})}{\left[\left(\frac{N}{2}-n\right)^2+\widetilde{P}\right] \Gamma(\frac{N}{2}-n-i \sqrt{\widetilde{P}}) \Gamma(\frac{N}{2}-n+i \sqrt{\widetilde{P}})},
\end{align} 
where $\Gamma(x)$, $\ceil{x}$  and $\floor{x}$ are the gamma, ceiling and floor functions respectively, and the first double sum has poles of order 2 while the second has simple poles. The canonical partition function is:
\begin{equation}
\begin{split}
	Z_c(N,\Delta,\widetilde{L})& =\delta _{\frac{N}{2}\left\lfloor \frac{N}{2}\right\rfloor } \sum _{n=0}^{\frac{N}{2}-1} \sum _{m=0}^n c_{nm\frac{N}{2}}\\
	+&\sum_{k=0}^{\left\lceil \frac{N}{2}\right\rceil-1 } \e^{-(\frac{N}{2} - k)^2 \widetilde{L}} \sum _{n=0}^k \sum _{m=0}^n (a_{nmk} \widetilde{L}+b_{nmk}) \\
		+&\sum _{k=0}^{\left\lfloor \frac{N}{2}\right\rfloor} \e^{-(N/2 - k)^2 \widetilde{L}} \sum _{n=\left\lceil \frac{N}{2}\right\rceil}^{N-k} \sum _{m=0}^{k} c_{nmk}\\
	+&\sum _{k=0}^{\left\lceil \frac{N}{2}\right\rceil -2} \e^{-(\frac{N}{2} - k)^2 \widetilde{L}} \sum _{n=k+1}^{\left\lceil \frac{N}{2}\right\rceil -1} \sum _{m=0}^k c_{nmk}, 
	\label{canon_z}
	\end{split}
\end{equation} 
where the coefficients $a_{ijk},b_{ijk}$ and $c_{ijk}$ are found in Appendix \ref{appx:familyofconstants}. Note that the term with the factor $\delta _{\frac{N}{2}\left\lfloor \frac{N}{2}\right\rfloor }$  vanishes for an odd number of counter-ions. 

	\subsubsection{Equation of state}	
	
		The pressure can be computed from the partition function or alternatively using the following contact relation (derived in Appendix \ref{sec:CT}): 
		\begin{equation}
			\widetilde{P}_c = -\frac{N^2}{4} + N\int_{-\infty}^{0}dx\; \widetilde{n}(x)  + \frac{2\Delta}{1+\Delta} \widetilde{n}(0). 
			\label{ConTeoNr_n}
		\end{equation}
		 This expression can be cast in terms of averages taken over all impermeable configurations each with weight $ z_{{N_{\ell},N_r}}(N,\Delta,\widetilde{L})/Z_c(N,\Delta,\widetilde{L})$:
		\begin{align}
			\widetilde{P}_c(N,\Delta,\widetilde{L}) = -\frac{N^2}{4} + N\left<N_{\ell}\right> + \frac{2\Delta}{1-\Delta} \left<N_{\ell}^2\right>,
			\label{ConTeoNr}
		\end{align}
		 which gives an explicit intuition of the effect of the dielectric jump: The sign of $\Delta$ determines if the third term of the equation is attractive or repulsive. As already seen in the one counter-ion case, $\Delta>0$ creates like-charges images  that increase the inter-colloid counter-ion density. The opposite behavior is seen with attractive image charges. The symmetry of the system allows to deduce limiting cases, as
		\begin{subequations}
			\begin{align}
				&\widetilde{P}_c(N,\Delta,0)=	\frac{N^2}{4} + \frac{2\Delta}{1-\Delta} \left<N_{\ell}^2\right>
				\label{N_P_lim_0},\\
				&\widetilde{P}_c(N,\Delta,\widetilde{L}) \xrightarrow[\widetilde{L} \to \infty]{}
				\begin{cases}
				-\frac{1}{4}  & N \text{ odd} \\
				\;\;\; 0 & N \text{ even}
				\end{cases}
				\label{N_P_lim_inf},
			\end{align}
		\end{subequations}
		through simple arguments. At zero length, the average number of particles on each side is $N/2$ by symmetry, which in equation \eqref{ConTeoNr} shows that for small separations, the effect of the dielectric jump is exclusively due to the fluctuation of counter-ions number. The term $\left<N_{\ell}^2\right>$ is straightforward to compute using the impermeable partition functions which are  given by $z_{{N_{\ell},N_r}}(N,\Delta,0) = [(1-\Delta)/(1+\Delta)]^N (N_r!N_{\ell}!)^{-2}$.  In the opposite limit, $L\to \infty$, the system tries to decouple into two neutral subsystems, one for each colloid. This process happens successfully for $N$ even, which leads to a vanishing attraction between the two neutral systems. This mechanism is frustrated when $N$ is odd, and thus cannot be exactly shared in two moieties. Then, the colloids are screened by $N-1$ counter-ions and together with the missing particle (delocalized in the interstitial region), they form an effective system of charge $-e/2$ which was already shown to have like-charge attraction of (rescaled) magnitude $-1/4$ when $L\to \infty$  \cite{tt2015}. The reason is that at such distances, the pressure becomes purely electrostatic, and the effective colloid of charge $-e/2$ on the 
		leftmost region, is attracted by the other rightmost effective colloid of charge $-e/2$, plus the delocalized ion of charge $e$. The net field on the leftmost colloid is thus attractive. These results are verified through the computation of the pressure from equation \eqref{canon_z}. 
	
		The isobaric pressure diverges as the colloids come close, which is expected to enforce zero length fluctuations. This marks a clear distinction between ensembles. In the opposite limit of infinite distance, the asymptotic behavior is the same as in the canonical ensemble:
		
		\begin{subequations}
			\begin{align}
			\label{zerolengthpressure_P}
				&\widetilde{P}(N,\Delta,\braket{\widetilde{L}}) \; \xrightarrow[\;\braket{\widetilde{L}} \to 0\;]{}  \; \infty,\\
			\label{inflengthpressure_P}	&\widetilde{P}(N,\Delta,\braket{\widetilde{L}})\xrightarrow[\braket{\widetilde{L}} \to \infty]{}
				\begin{cases}
					-\frac{1}{4}  & N \text{ odd} \\
					\;\;\;0 & N \text{ even}.
				\end{cases}
			\end{align}
		\end{subequations}
	
\begin{figure}[ht]
		\centering	        	\includegraphics[width=0.48\textwidth]{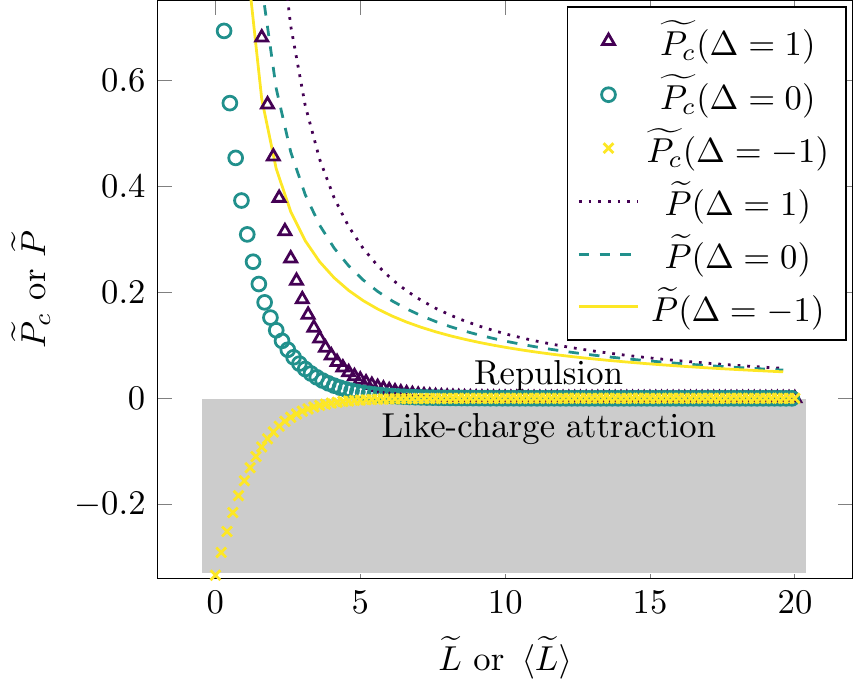}%
			\caption{
				Equation of state for the canonical (symbols) and isobaric (lines) ensemble with permeable colloids and $N=2$ counter-ions. The pressure vanishes asymptotically for large distances (see Eqs. \eqref{N_P_lim_inf},\eqref{inflengthpressure_P}), while the zero length pressure depends on the ensemble and dielectric jump (see Eqs. \eqref{N_P_lim_0}, \eqref{zerolengthpressure_P}). The possibility for like-charge attraction in a fixed length system exists for $\Delta< -3/5$ (plot inside gray shade) with a minimum value of $\widetilde{P}_c = -1/3$. Note that the only canonical pressure that diverges at $\widetilde{L} = 0$ is that for $\Delta = 1$, as was the case for $N=1$.}
			\label{fig:P_2}
	\end{figure}

\begin{figure}[ht]
		\centering                    \includegraphics[width=0.48\textwidth]{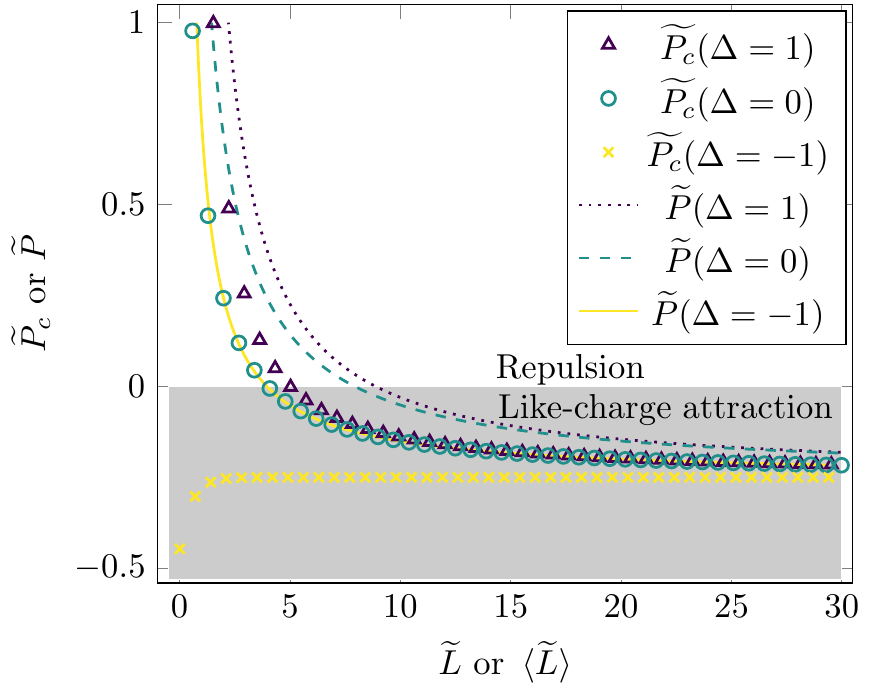}%
			\caption{
				Equation of state for the canonical (symbols) and isobaric (lines) ensemble with permeable colloids and $N=3$ counter-ions. Note the coincidental proximity of $\widetilde{P}_c(\Delta = 0)$ and $\widetilde{P}(\Delta = -1)$. Yet, while $\widetilde{P}_c(\Delta = 0)$ has a finite value when $\widetilde{L} = 0$, $\widetilde{P}(\Delta = -1)$ diverges as $ \braket{\widetilde{L}} \to 0$}
			\label{fig:P_3}
	\end{figure}

	Both the canonical and isobaric ensembles share a dichotomy of the pressure 
	behavior,  depending on the parity of $N$. In this sense, the qualitative behavior is summarized for both ensembles by the cases $N=1$ (Fig.\ref{fig:P_1}) and $N=2$ (Fig.\ref{fig:P_2}). Take for example $N=3$ (Fig.\ref{fig:P_3}): although the equation of state cannot be quantitatively described by the $N=1$ case, the key features such as asymptotic values, presence of like-charge attraction and $\Delta$ dependence are the same. The same is true for any even $N$ and $N = 2$.         

\subsubsection{Counter-ion density}	
\label{sec:counterion_density}

	 We have seen that there is an explicit connection between pressure, density and average number of particles in each region. In this section, the counter-ion density profile is computed explicitly.
	 Specifically, the density at the dielectric jump is found to be proportional to the average  squared outside number of counter-ions. Besides, the aforementioned separation in effective objects at large separation has an {explicit} 
	 fingerprint on the density profile.
	 
	 	\begin{figure}[htb]
	 	\centering	
		\includegraphics[width=0.48\textwidth]{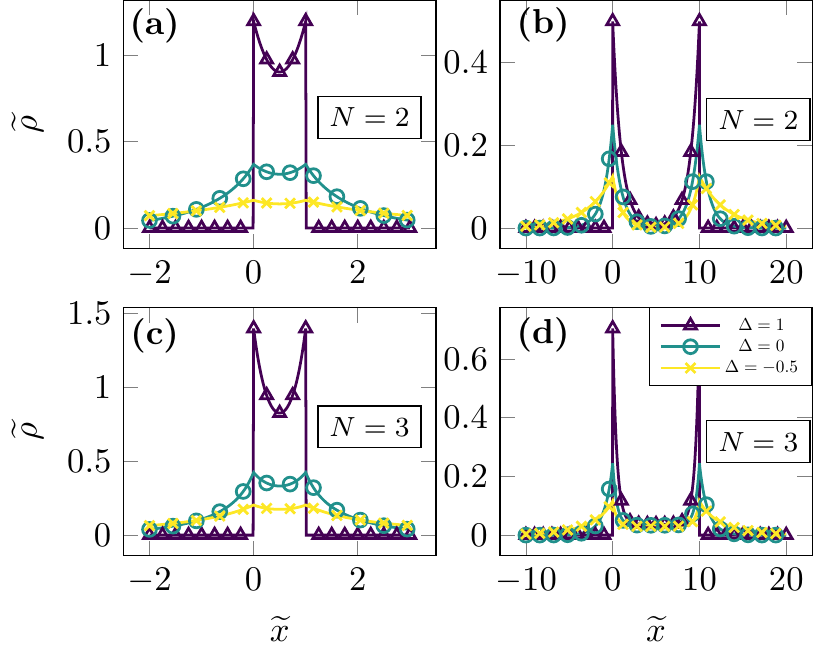}%
	 		\caption{
	 			Normalized counterion density ($\widetilde{\rho} = \widetilde{n}/N$) for permeable colloids at distances $\widetilde{L} = 1$ (left column) and 10 (right column), with $N=2$ counterions (top row) or 3 counter-ions (bottom row). 
	 			For all cases, the counter-ions are expelled to the exterior regions as the dielectric jump $\Delta$ goes to $-1$, and conversely drawn inside in the opposite limit $\Delta \to 1$. For large separations, the even case (panel b) decouples with a nearly vanishing density at $\widetilde{L}/2$. Instead, the odd case (panel d) shows an almost constant density in the middle region, which accounts for the counter-ion that is shared between the colloids: this causes colloids to attract each other. }
	 		\label{fig:densityN}
	 \end{figure}
	 
	 The density profile $\widetilde{n}(\widetilde{x},\Delta,N,\widetilde{L})$ is defined piece-wise for computational convenience just as when $N=1$: $\widetilde{n}(x<0),\widetilde{n}(0<\widetilde{x}<\widetilde{L})$ and $\widetilde{n}(\widetilde{x}>\widetilde{L})$ for the left, middle and right regions respectively. 
	By symmetry, $\widetilde{x}>L$ is computed using $\widetilde{n}(\widetilde{x}) = \widetilde{n}(\widetilde{L}-\widetilde{x})$. The results are expressed in terms of the partition functions and density profiles  obtained for the case with fixed number of counterions per region which can be found in \cite{vtt2016}: 	
		\begin{align}
			\widetilde{n}(\widetilde{x}) = \sum_{N_{\ell} = 0}^{N}\sum_{N_{r} = 0}^{N-N_{\ell}} \frac{z_{N_{\ell},N_r}(N,\Delta,\widetilde{L}) }{Z(N,\Delta,\widetilde{L})} {\widetilde{n}}_{N_{\ell},N_r}(\widetilde{x},N,\Delta,\widetilde{L})\label{N_n},
		\end{align}
	where  ${\widetilde{n}}_{N_{\ell},N_r}$ is the density profile for an impermeable colloid (Eq. \eqref{imperm_n1} and \eqref{imperm_n2}). The normalized counterion density $\widetilde{\rho} = \widetilde{n}/N$, displayed in Fig. \ref{fig:densityN}, behaves as already observed with $N=1$: $\Delta$, which is associated to the sign of the image charges, regulates the population of counter-ions in each region while the dimensionless length determines if the system has decoupled into two screened colloids depending on whether their double layers decouple or not, according to the parity of $N$. 
	The situations with $N=1$ and $N=2$ turn out to be emblematic of the odd and even $N$ cases, respectively
	(see Figs \ref{fig:densityN1} and \ref{fig:densityN}).

	From equation \eqref{CT_N_2} we know that ${\widetilde{n}}_{N_{\ell},N_r}(0^{-}) = (1+\Delta)/(1-\Delta)N_{\ell}^2$ and thus $\widetilde{n}(0)$ is proportional to $\braket{N_{\ell}^2}$. Therefore, the pressure (Eq. \eqref{ConTeoNr}) can be cast either in terms of $\widetilde{n}(0)$ or $\braket{N_{\ell}^2}$. We can obtain compact results for the infinite length limit of those two moments. The following expressions are for $N>1$:
	\begin{subequations}
		\label{Nl_moments}			
		\begin{align}
			\lim\limits_{\widetilde{L}\to \infty} \braket{N_{\ell}} = 
			\begin{dcases}
			\frac{\sum _{n=0}^p \sum _{m=0}^n \frac{m+n}{2}  a_{nmp}}{\sum _{n=0}^p \sum _{m=0}^n a_{nmp}} & N=2p+1\\
			\frac{\sum _{n=0}^p \sum _{m=0}^n \frac{m+n}{2}  c_{nmp}}{\sum _{n=0}^p \sum _{m=0}^n c_{nmp}} & N = 2p,
			\end{dcases}\\
			\lim\limits_{\widetilde{L}\to \infty} \braket{N_{\ell}^2} =
			\begin{dcases}
			\frac{\sum _{n=0}^p \sum _{m=0}^n \frac{m^2+n^2}{2}  a_{nmp}}{\sum _{n=0}^p \sum _{m=0}^n a_{nmp}} & N =2p+1\\
			\frac{\sum _{n=0}^p \sum _{m=0}^n \frac{m^2+n^2}{2}  c_{nmp}}{\sum _{n=0}^p \sum _{m=0}^n c_{nmp}} & N = 2p.
			\end{dcases}
		\end{align}
	\end{subequations}

	When $\Delta = 0$, the system tries to decouple into two symmetric neutral subsystems, succeeding when $N$ is even and failing otherwise, with each screened colloid having half of its counterions on each side. This allows to have a very good estimate of the average number of left particles  $\braket{N_{\ell}} \approx \floor{N/2}/2 = p/2 $ when $\widetilde{L} \to \infty$ and $ N = 2p+1 $ or $2p$ for each parity case respectively. In the limiting cases $\Delta = 1$ we have $\braket{N_{\ell}} = 0$ and in the opposite case $\Delta = -1$ we get $\braket{N_{\ell}} = p$. These results are generalized in Table \ref{tab:mean}.

		\begin{table}[ht]
			\centering
				\caption{Asymptotic behavior of the mean outside particle number}
				\label{tab:momentsasymp}
				\begin{tabular}{l|l|l} 
					$N$ & $\lim\limits_{\widetilde{L}\to \infty}$ 	$\braket{N_{\ell}}$& $\lim\limits_{\widetilde{L}\to \infty}$ 	$\braket{N_{\ell}^2}$\\
					\hline
					1 & $0$ & $0$\\
					2 & $\frac{1}{2}-\frac{\Delta }{2}$ & $\frac{1}{2}-\frac{\Delta }{2}$\\
					3 & $2-\frac{4}{3-\Delta}$ & $2-\frac{4}{3-\Delta}$\\	
					4 & $1+\frac{2 \Delta }{\Delta ^2-3}$& $\frac{4 (1-\Delta)}{3-\Delta ^2}$\\		
					5 & $\frac{6 (1-\Delta)}{5-\Delta ^2-2 \Delta}$& $\frac{6 (4-3 \Delta )}{5-\Delta ^2-2 \Delta}-3$\\	
					6 & $\frac{3}{2}-\frac{2 \Delta }{5-3 \Delta ^2}-\frac{\Delta }{2} $& $\frac{3}{2}+\frac{6 (1-\Delta)}{5-3 \Delta ^2}-\frac{3 \Delta }{2} $\\	
				\end{tabular}
			\label{tab:mean}
		\end{table}
	
	\subsubsection{Fluctuations}
	
\begin{figure}[ht]
			\centering	
		    \includegraphics[width=0.48\textwidth]{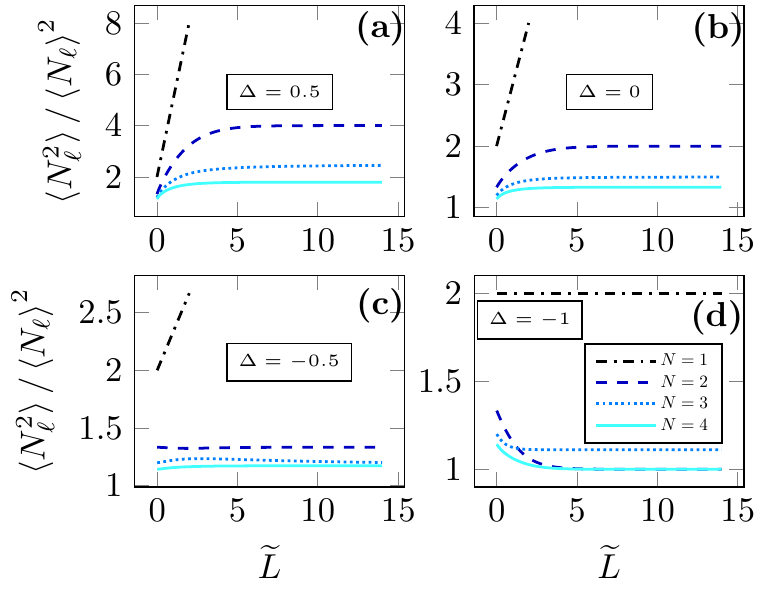}%
				\caption{Fluctuations for $N\leq 4$ counter-ions and different values of the $\Delta$ jumps: (a) $\Delta = 0.5$, (b) $\Delta = 0$, (c) $\Delta = -0.5$ and (d) $\Delta = -1$. For $N>1$, the fluctuations saturate for a large enough $\widetilde{L}$.}
				\label{fig:fluctuations}
		\end{figure}

		We now proceed to examine the fluctuations of the left side number of particles, which are the same for the right side. These fluctuations are an increasing function of $\Delta$ (except for $\widetilde{L}=0$ where $\Delta$ is irrelevant). This can be seen in terms of the positive image charges that drive the counterions close to the colloid (see the particle densities in figures \ref{fig:densityN1} and \ref{fig:densityN}) which favors the ``crossing'' of counterions between interior and exterior regions.
		The fluctuations ${\braket{N_{\ell}^2}}/{\braket{N_{\ell}}^2}$ follow from the previously defined moments for the number of left counter-ions, see Fig. \ref{fig:fluctuations}.
		
		From the limiting behavior found in equations \eqref{Nl_moments}, we can extract the asymptotic value for $\widetilde{L}\to \infty$ (except when $N=1$ which has an oblique asymptote, see Eq. \eqref{1i-fluc}).  Notice that as the number of counter-ions increases, the characteristic separation length for which the fluctuations reach their terminal value approaches 0. 
		

\section{When does mean-field apply?}
\label{sec:mf}

	We finally address the connection to mean-field results,	where Poisson's equation ruling the behavior of the mean electrostatic potential $\phi(x)$ is closed by the assumption that the ionic density is given by the Boltzmann distribution:  $n_{\text{\tiny PB}}(x)=n_0 \exp(-e\beta \phi)$, $n_0$ being some normalization constant. This results into the Poisson-Boltzmann equation
	(PB):
	\begin{equation}
		{\phi}''({x}) = -\frac{2en_0}{\epsilon({x})} \e^{-e\beta{\phi}({x})}. \label{PB}
	\end{equation}
	where the factor $2$ stems from the Poisson equation convention \cite{poissoneq}.
	
	We expect this framework to become operational under conditions of weak electrostatic coupling: the colloid charge $q$ being fixed, this is achieved when 
	$e\to 0$, while of course keeping the electroneutrality constraint satisfied ($Ne=2q$, meaning that $N\to \infty$). The rescaled length used in previous sections, involving the charge $e$, becomes inadequate and has to be slightly modified. Poisson-Boltzmann
    equation is solved piece-wise and the solutions are matched with the continuity of $\phi$ and the discontinuity of the electric field $\phi'$ due to the fixed charges. The pressure $P_{\text{\tiny PB}}$ follows from the contact theorem, which, quite remarkably, also holds within mean-field. Introducing the rescaled pressure $\widehat{P}_{\text{\tiny PB}} =P_{\text{\tiny PB}} (\epsilon_1+\epsilon_2) /(2q^2)$, we get
    \begin{equation}
		\sqrt{1-\Delta} \sec\left(\frac{\widehat{P}_{\text{\tiny PB}}^{\frac{1}{2}}\widehat{L}}{\sqrt{1+\Delta}} \right) +\sqrt{1+\Delta} \tan\left(\frac{\widehat{P}_{\text{\tiny PB}}^{\frac{1}{2}}\widehat{L}}{\sqrt{1+\Delta}}\right) =  \frac{1}{\widehat{P}_{\text{\tiny PB}}^{\frac{1}{2}}} 
		\label{po2}
    \end{equation}
    where we have introduced the rescaled length $\widehat{L} = L\beta qe/(\epsilon_1+\epsilon_2) $.

	\begin{figure}[ht]
	    \centering
		\includegraphics[width=0.48\textwidth]{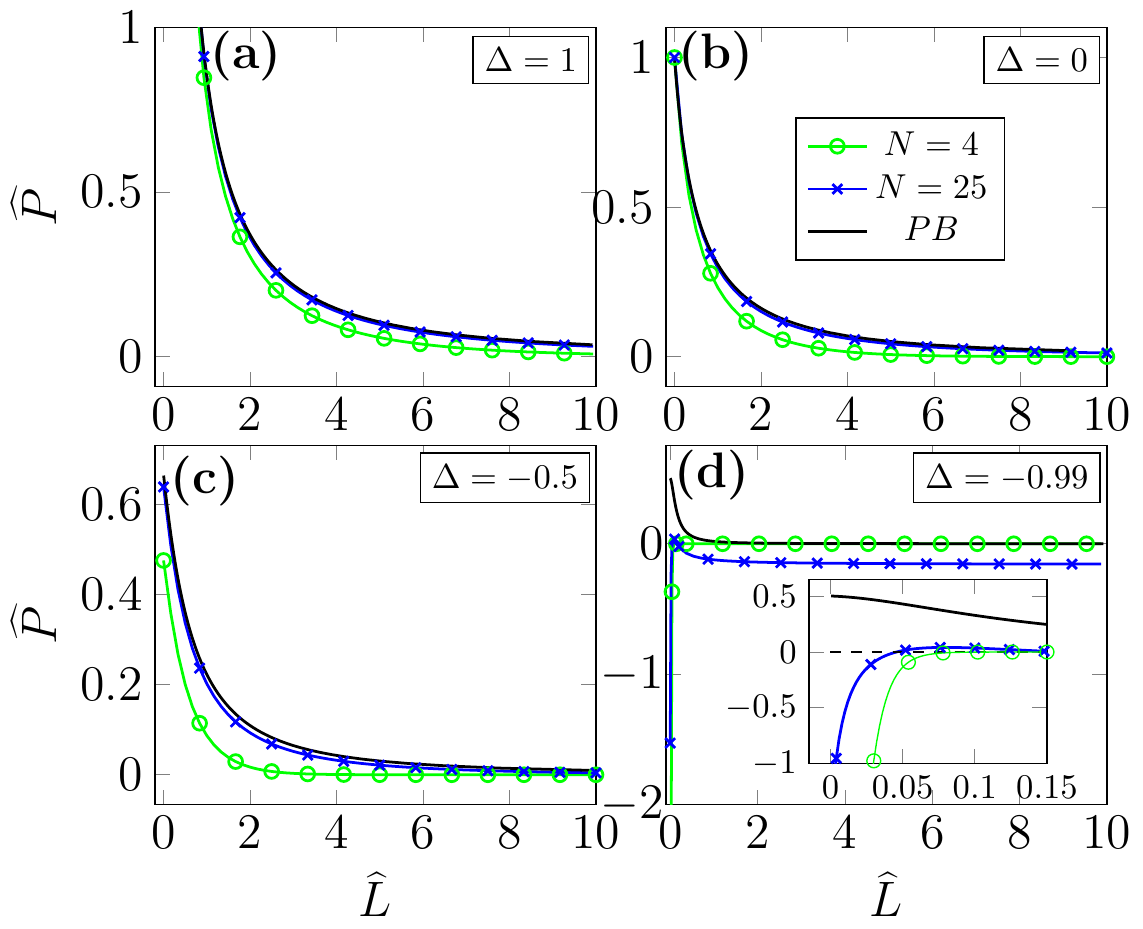}%
		\caption{
		Comparison of the Poisson-Boltzmann $\widehat{P}_{\text{\tiny PB}}$ and exact pressures for $N = 4$ and $25$. The colloids charge $q$ is fixed and the counter-ions charge varies, for the exact calculation, as $e = q/2N$. In the limit $N\to \infty$, $e \to 0$ while keeping $eN=2q$ fixed, mean-field theory becomes exact. Note that with as few as 25 counter-ions, the Poisson-Boltzmann pressure yields good results (panels b-c), except for $\Delta = -0.99$ (panel a) where the coupling constant \eqref{xi} is very large. The inset zooms the small $\widehat{L}$ values for which PB is quantitatively and qualitatively off with respect to the exact calculation. As $\Delta$ approaches $-1$, more counterions are required to be in the weak-coupling regime $\Xi_{\text{in}} \ll 1$. Note that like-charge attraction is completely lost in the Poisson-Boltzmann theory \cite{Neu1999}.}
		\label{fig:PBP}
	\end{figure}
	
	For $\Delta = 1$, we recover the ``impermeable'' results with all counterions in the interstitial region, obtained in \cite{tt2015} by taking the mean-field limit from an exact description, see also \cite{Kanduc2008}. Besides the pressure,  it is interesting to see how the ionic density within the exact treatment compares with the mean-field limit. Introducing $\widehat{n}_{\text{\tiny PB}} = n_{\text{\tiny PB}}(\epsilon_1+\epsilon_2)/q^2 \beta$, the rescaled density profile is: 
	\begin{align}
	\frac{\widehat{n}_{\text{\tiny PB}}(\widehat{x})}{\widehat{P}_{\text{\tiny PB}}}	 = \begin{cases}
		\sec^2\left(\frac{(2\widehat{x}-\widetilde{L})\widehat{P}_{\text{\tiny PB}}^{\frac{1}{2}} }{\sqrt{1+\Delta}} \right),	& \widehat{x}  \in [0,\widetilde{L}] \\
		\left[\frac{2\left(\widehat{x} - \widehat{L}\right)\widehat{P}_{\text{\tiny PB}}^{\frac{1}{2}}}{\sqrt{1-\Delta}}  + \cos\left(\frac{\widehat{P}_{\text{\tiny PB}}^{\frac{1}{2}}\widehat{L}}{\sqrt{1+\Delta}} \right)  \right]^{-2},
		& \widehat{x} \not \in [0,\widetilde{L}] 
		\end{cases}
	\end{align}

    \begin{figure}[ht]
	    \centering	
		\includegraphics[width=0.48\textwidth]{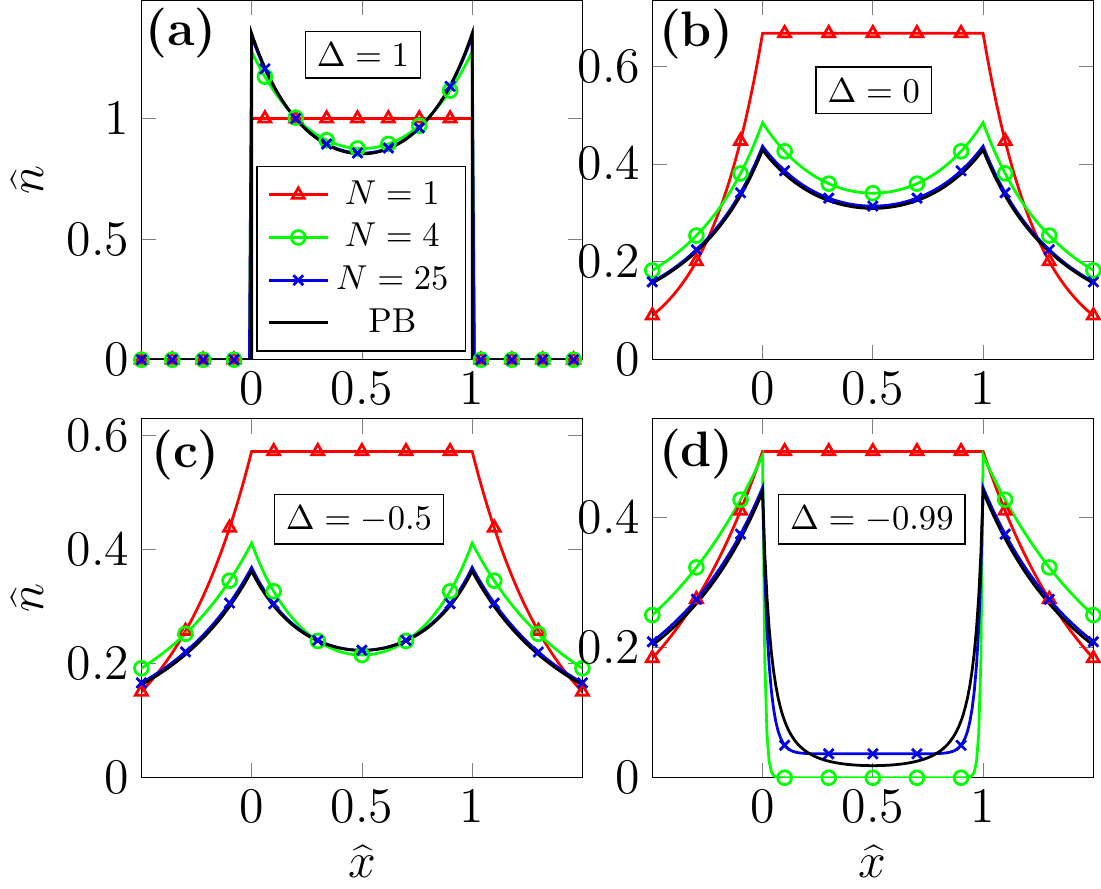}%
		\caption{
			Density profile for $N=1,4,25$ and $\widehat{L} = 1$: comparison of the exact results with the 
			Poisson-Boltzmann approximation, $\widehat{n}_{\text{\tiny PB}}$. For $N=25$ counterions, the mean-field theory yields a fair approximation. The panels show a range of dielectric jumps: (a) $\Delta = 1$, (b) $\Delta = 0$, (c) $\Delta = -0.5$ and (d) $\Delta = -0.99$.}
		\label{fig:PBn}
    \end{figure}

	The results for the pressures and the density profiles are shown in Figs. \ref{fig:PBP} and \ref{fig:PBn}.
	They reveal that for as few as $N=4$ counter-ions, the mean-field approach gives a good approximation to exact results, specially for non-negative dielectric jumps. To understand why this is happening, we
	examine the coupling constants:   $\Xi_{\text{out}}$ and $\Xi_{\text{in}}$ for the outside ($\epsilon_1$, left and right regions) and  inner sectors ($\epsilon_2$, middle region) respectively. These quantities follow from 
	comparing the ion-ion typical interaction energy, discarded at PB level, to $kT$:		\begin{equation}
	    \Xi_\alpha \,=\, \frac{e^2 a_\alpha/\epsilon_\alpha}{kT}  , \quad
	    \alpha=\text{in, out}
	\end{equation}
	where the numerator is the 
	the typical electrostatic work  needed to compress a pair of counterions in the system, and we take $a_{\alpha} = kT\epsilon_{\alpha}/(e^2\braket{N_{\alpha}}/2)$ as the average  counterion separation in the region of interest. This length follows from the quotient of the double-layer length in each region and the corresponding number of counterions there. The double-layer length was computed exactly for the impermeable case with all counterions in between the colloids \cite{tt2015}. Its size is of order $(kT\epsilon_{\text{in}}/e^2) (N-1)/(N+1)$ for any colloid separation $\widetilde{L}$, and therefore $N$ independent whenever $N$ exceeds a 
	few units. Then, for the impermeable case worked out in 
	\cite{tt2015}, the average length between counterions behaves like $ kT\epsilon_{\text{in}}/(e^2(N/2))$, where $N/2$ is the counterions in each double-layer. It can be checked that this result generalizes to a permeable system by replacing the corresponding number of counterions for the double-layer at each region $N/2 \to \braket{N_{\alpha}}/2$ and using the respective dielectric constant.  
	
	The exact expressions of $\braket{N_{\text{out}}}$ and $\braket{N_{\text{in}}}$ are cumbersome, but we are only interested in their limiting behavior as $\Delta \to \pm 1$,
	corresponding to $\epsilon_1 \ll \epsilon_2$, or to the reverse. These limits lead to a depletion in a given region (a small $\braket{N_\alpha}$), which entails the
    failure of mean-field. We then proceed to estimating $\braket{N_{\alpha}}$  using the results reported in section \ref{sec:counterion_density}. We focus on large enough $\widetilde{L}$.
    For $\braket{N_{\text{out}}}$ we know that as $\Delta \to -1$ it approaches $N/2$ and as $\Delta \to 1$ it goes to zero as $\braket{N_{\text{out}}} \sim (1-\Delta)/(1+\Delta)$ (this follows from Eq. \eqref{Nl_moments} and $\braket{N_{\text{out}}} = 2 \braket{N_{\ell}}$). We can condense both behaviors using $\braket{N_{\text{out}}} \sim N(1-\Delta)/2$.  In a similar fashion, $\braket{N_{\text{in}}} \sim N(1+\Delta)/2$ and thus the coupling constants are defined as:
 	\begin{align}
 	\label{xiout}
		\Xi_{\text{out}} 
		\,= \frac{4}{N(1-\Delta)}, \\
	    \Xi_{\text{in}} \,= \frac{4}{N(1+\Delta)} 
		.
		\label{xi}
	\end{align}

    \begin{figure}[ht]
	    \centering	
		\includegraphics[width=0.48\textwidth]{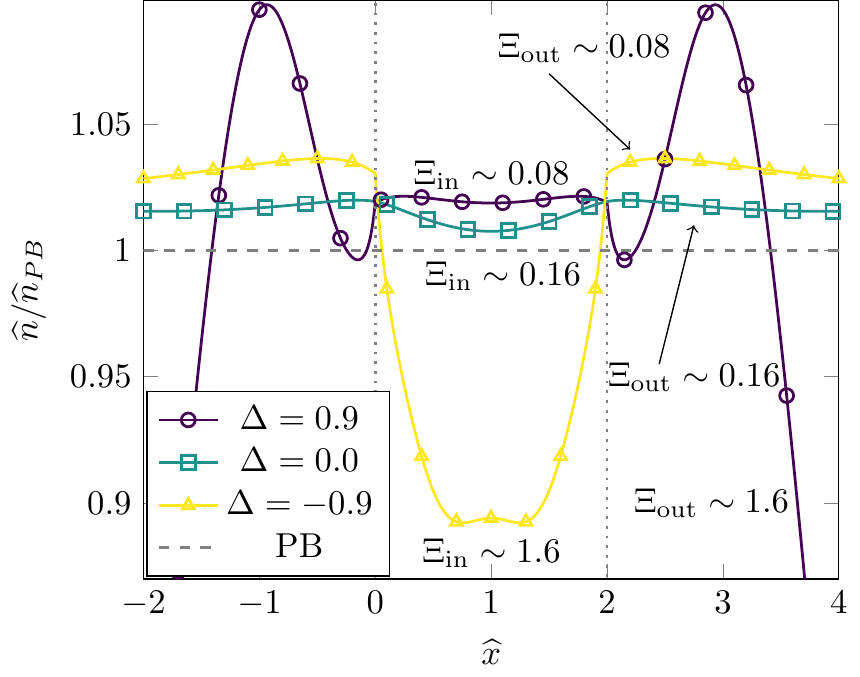}%
		\caption{
			Density profile $\widehat{n}$ for $N = 25$ and $\widehat{L} = 2$ normalized by Poisson-Boltzmann solution  $\widehat{n}_{PB}$. The colloids positions are marked by the two vertical dotted lines
			delimiting the ``in'' region, and the horizontal dashed line marks the mean-field behavior.
			Such a plot is more appropriate than Fig. \ref{fig:PBn} to appreciate the mean-field departure 
			from the exact solution,  depending on the region (in or out). Note that as $\Delta$ decreases and approaches $-1$, the quotient $\widehat{n}/\widehat{n}_{PB}$ departs further and further from unity in the inner region: the Poisson-Boltzmann solution is then less and less accurate, as embodied in the value of $\Xi_{\text{in}}$  (see Eq. \eqref{xi}).}
		\label{fig:PBnin}
    \end{figure}
    
    \begin{figure}[ht]
	    \centering	
		\includegraphics[width=0.48\textwidth]{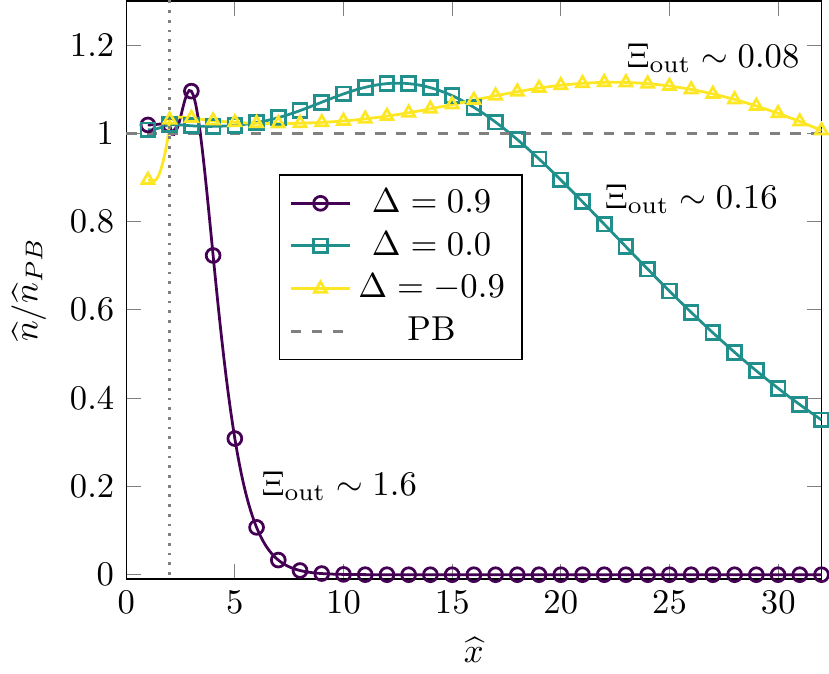}%
		\caption{Same as Fig. \ref{fig:PBnin},           focusing onto the right side of the exterior region. 
			Note that as $\Delta$ increases and approaches 1, 
			mean-field departs further and further from the exact density, 
			as embodied in Eq. \eqref{xiout}. 
			}
		\label{fig:PBnout}
    \end{figure}    	

	 The weak-coupling regime, defined by $\Xi_{\alpha} \ll 1$ ($\alpha=\text{in, out}$), is suitable for a mean-field description, when the discrete nature of counterions can be neglected; for $\Delta=0$, we recover the results reported in \cite{Dean2009,tt2015}. This regime is met when $N$ increases beyond a few units, irrespective of temperature. The irrelevance of temperature is specific to one-dimensional systems. In dimensions 2 and 3, increasing $T$ 
	 leads to a decrease of $\Xi$, bringing closer to the mean-field regime.
	 We see in Figs. \ref{fig:PBP} and \ref{fig:PBn} that the Poisson-Boltzmann theory gives a good approximation of the exact system when $N = 25$. 
	 There is an exception when $\Delta = -1$ (Fig. \ref{fig:PBP}) for which the mean-field pressure vanishes while its exact counterpart has negative values.  This is expected due to the term $1/(1+\Delta)$ in the coupling parameter; as $\Delta$ approaches $-1$, a greater number of counterions is required to be in the mean-field regime. The effect of $\Delta$ in each regions is better seen in Figs. \ref{fig:PBnin}-\ref{fig:PBnout}, where the local density profile $\widehat{n}(\widehat{x})$ with $N=25$ is compared to the Poisson-Boltzmann solution $\widehat{n}_{\text{PB}}(\widehat{x})$. 
	 These deviations are seen to increase in each region according to the respective coupling constant.
	 Note that $\Xi_{\alpha}$ is inversely proportional to the number of counterions, which is the same behavior found for an impermeable colloid with all the counterions in the middle region \cite{tt2015,Dean2009}.


\section{Conclusions}

We have obtained the exact solution for a schematic one-dimensional colloidal model with an arbitrary number of counter-ions, in the presence of dielectric discontinuities (see Fig. \ref{sketch}). The colloids are either impermeable or not to the counterions. We find that the pressure, which in 1D coincides with the force, can assume negative values (see Figs. \ref{fig:P_1}, \ref{fig:P_2}); there is like-charge attraction in a given domain determined by the distance between colloidal particles, the dielectric jump and the number of counterions. Unlike for a uniform dielectric medium, the presence of a dielectric discontinuity enables the possibility for like-charge attraction in a permeable colloid for any $N$, regardless of its parity. 
Additionally, we find a contact theorem-like relationship that connects density to pressure. This allows to see how the image charges, induced by the dielectric discontinuity, shape the counterion density through attraction or repulsion and thus the interaction among colloids.
Both the pressure (Fig.  \ref{fig:PBP}) and density profile (Fig.  \ref{fig:PBn}) are shown to converge towards the mean-field prediction: for a large number of counterions, the Poisson-Boltzmann equation is in excellent agreement with the exact theory. This is consistent with 1D strong coupling parameter found in previous works \cite{Dean2009,tt2015}: besides the total number of ions, the validity of mean-field here depends on the dielectric discontinuity, but not on temperature, at variance with two or three dimensional systems.

\section*{Acknowledgement}

We would like to thank I. Palaia and L. \v{S}amaj for useful discussions.
This work was supported by an ECOS-Nord/Minciencias C18P01 action of Colombian and French cooperation. L.V and G.T. acknowledge support from Fondo de Investigaciones, Facultad de Ciencias, Universidad de los Andes INV-2019-84-1825 and Exacore HPC Uniandes for providing high performance computing time. L.V acknowledges support from Action Doctorale Internationale (ADI 2018) de l'IDEX Universit\'e Paris-Saclay.


\appendix

\section{Potential from dielectric images}
\label{appendix:potential}

\begin{figure}[ht]
	\centering
	\includegraphics[width=0.48\textwidth]{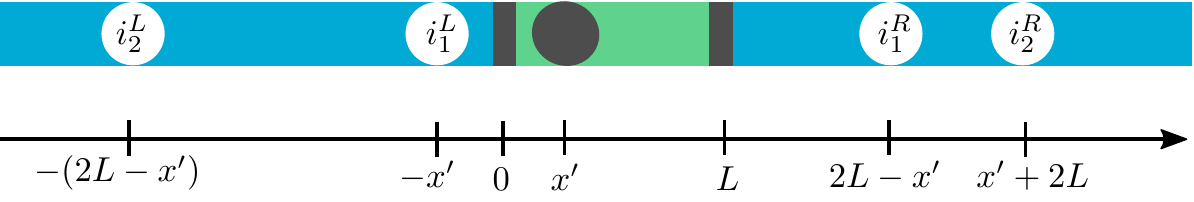}
	\caption{
		 Sketch of the image charge construction to compute the potential $V$ due to the dielectric discontinuities. The super-indices L and R are for images generated by reflection upon the left (at $x=0$) or right (at $x=L$) boundaries respectively. The sub-index indicates the generation of the image, where $1$ is created by the central counter-ion itself, and the following generations are created by the images of the images etc., in an iterative manner.  
	}
	\label{fig:imagecharges}	
\end{figure} 

In this appendix, we provide the expression of the 1D potential $V(x,x')$, created at $x$ by a charge $q$ at $x'$;
it is a solution of the Poisson equation \cite{poissoneq}.
The particle is in a piece-wise linear dielectric medium defined by $\epsilon_1$ and $\epsilon_2$ for ${x}\not\in[0,{L}]$ and ${x}\in[0,{L}]$ respectively, as seen in Fig. \ref{fig:imagecharges}. $V$ can be computed using the method of images as follows.

Consider a charge $q$ at $0<x'<L$ (Fig.\ref{fig:imagecharges}). This charge creates a series of images $i_k^L$ and $i_k^R$ with respect to the left and right colloids. Their respective positions are $x^L_k$ and $x^R_k$:
	\begin{align*}
	x^L_1 &= -x',  & x^R_1 &= 2L-x' \\
	x^L_2 &= -(2L-x'),  & x^R_2 &=2L+x' \\
	x^L_3 &= -(2L+x'),  & x^R_3 &= 4L-x'\\
	x^L_4 &= -(4L-x'),  & x^R_4 &=4L +x'\\
	 &\;\;\vdots &  &\;\;\vdots
	\end{align*}		
Each image creates a potential $V_k^{\alpha}(x) = -q\Delta^k|x-x^{\alpha}_k|/\epsilon_2$ ($\alpha = L,R$) at $x\in[0,L]$. Therefore, the potential created by the image charges is:

	\begin{subequations}
	\begin{align}
	\sum_{k=1}^{\infty} (V_k^L + V_k^R) 	&= -\frac{2Lq}{\epsilon_2}\sum_{k=1}^{\infty}k\Delta^k\\
	& = -\frac{2Lq\Delta }{\epsilon_2(1-\Delta)^2}
	\end{align}
	\end{subequations}
Finally the potential for $x,x'\in[0,L]$ is:
\begin{subequations}
\begin{align}
V(x',x) &= -\frac{q}{\epsilon_2}|x'-x|+\sum_{k=1}^{\infty} (V_k^L + V_k^R)\\
&= -\frac{q}{\epsilon_2}|x'-x| -\frac{2L\Delta }{\epsilon_2(1-\Delta)^2}\label{V1}
\end{align}
\end{subequations}

Now for $x<0$, we use a solution to Poisson equation where the is no charge:
\begin{equation}
V(x<0,x') = 
ax + b .
\label{V2}
\end{equation}
The constants $a$ and $b$ are found using the continuity of $V$ and the displacement field at $x=0$ . By enforcing these conditions on Eqs. \eqref{V2} and \eqref{V1} it follows that:
\begin{align}
\epsilon_1 a &= q \\
 b &= -\frac{qx'}{\epsilon_2} -\frac{2L\Delta }{\epsilon_2(1-\Delta)^2}
\end{align}
and therefore
\begin{equation}
V(x<0,x') = 
\frac{qx}{\epsilon_1} - \frac{qx'}{\epsilon_2} -\frac{2L\Delta }{\epsilon_2(1-\Delta)^2} .
\end{equation}

The missing $x>L$ case is found using an analog procedure applied to the continuity at $x =L$. Finally, the situations with $x'<0$ and $x'>L$ can also be found using the symmetry of the potential:
\begin{subequations}
\begin{align}
V(0<x<L,x'<0) &= V(x'<0,0<x<L) 
\\&= 	\frac{qx'}{\epsilon_1} - \frac{qx}{\epsilon_2} -\frac{2L\Delta }{\epsilon_2(1-\Delta)^2}
\label{V3}
\end{align}
\end{subequations}
By solving again the Poisson equation (but now in a region with charge), we have:
\begin{equation}
V(x<0,x'<0) = 
-\frac{q|x'-x|}{\epsilon_1} + b_2
\label{V4}
\end{equation}
and $b_2$ follows from comparing \eqref{V3} and \eqref{V4}:
\begin{equation}
\frac{qx'}{\epsilon_1} + b_2 = \frac{qx'}{\epsilon_1}  -\frac{2L\Delta }{\epsilon_2(1-\Delta)^2}
\end{equation}
and finally:
\begin{equation}
V(x<0,x'<0) = 
-\frac{q|x'-x|}{\epsilon_1}  -\frac{2L\Delta }{\epsilon_2(1-\Delta)^2}.
\end{equation}

To summarize, the potential created at point $x$ by a charge $q$ located at $x'$ is defined piece-wise as: \begin{equation}
	V({x},x'<0) = 
	\begin{cases}
	-\frac{q \left| {x}-x'\right| }{\epsilon_1}-  \frac{2L q\Delta}{(1-\Delta )^2 \epsilon_2},  & {x} < 0 \\
	-\frac{q {x}}{\epsilon_2}+\frac{q x'}{ \epsilon_1} - \frac{2L q\Delta }{(1-\Delta )^2 \epsilon_2}, & 0< {x} < {L} \\
	-\frac{q {x}}{ \epsilon_1}+\frac{q x'}{ \epsilon_1}  - \frac{2L q\Delta ^2 }{(1-\Delta )^2 \epsilon_2}, & {x} > {L}, 
	\end{cases}
	\label{potl}
\end{equation}

\begin{equation}
	V({x},0<x'<L) =
	\begin{cases}
	\frac{q {x}}{\epsilon_1}-\frac{q x'}{\epsilon_2}  - \frac{2L q\Delta}{(1-\Delta )^2 \epsilon_2}, & {x} < 0 \\
	-\frac{q \left| {x}-x'\right| }{ \epsilon_2}- \frac{2L q\Delta}{(1-\Delta )^2 \epsilon_2}, & 0< {x} < {L} \\
	-\frac{q {x}}{\epsilon_1}+\frac{q x'}{\epsilon_2} - \frac{2L q\Delta^2 }{(1-\Delta )^2 \epsilon_2}, & {x} > {L}.
	\end{cases}
	\label{potm}
\end{equation}
 The missing right potential is computed by symmetry using $V(x,x'>L) = V(L-x,L-x')$. 
 An alternative derivation follows from integrating the transverse degrees of freedom of the equivalent 3D system (see \cite{Wang2019} for a multilayered dielectric medium).  

 The information conveyed by these relations is that when a charge $q$ sits at $x'$, the modulus of the electric field created at point $x$ is always $q/\epsilon_i$ if $x$ is in the region with permittivity $\epsilon_i$, while the direction of the force changes depending if $x$ is located on the right, or on the left of the source at $x'$.
	 
\section{Overlapping equations of state}
\label{appx:deltaequivalance}

This appendix contains details on the mapping between
the canonical ensemble with $\Delta = 1$ and the isobaric ensemble with $\Delta = -1$, when $N=1$. Figure \ref{fig:P_1} illustrates this correspondence.
We have shown in the main text that the equation of state reads, in both cases:
\begin{equation}
    \widetilde{P}_c \,=\, \widetilde P \,=\,  \frac{1}{\widetilde{L}} -\frac{1}{4},
\label{eq:app_eos1}
\end{equation}
at the expense of a slight abuse of notation (replacing $\widetilde{L}$ by $\braket{\widetilde{L}}$ in the isobaric case).
When $\Delta =1$, the counterion is confined between the colloids, unlike when $\Delta = -1$ where it is expelled from the middle region. The confined case can be understood intuitively: the repulsive term, $1/\widetilde L$ in \eqref{eq:app_eos1},
is the density of an ideal gas in a box of size $L$; the counterion indeed is in a zero-field region. The attractive term is the electric force exerted on the right colloid by the other colloid and the counterion.

On the other hand, the case $\Delta = -1$ features an empty middle region; the attractive image charges force the counterion to be either on the left or right region. By symmetry, the two possibilities have the same pressure contribution. To understand the isobaric result at $\Delta=-1$, we consider that the right colloid is now able to fluctuate in position, while the left one is fixed. An external operator exerts a force $P$ onto the right colloid (with the convention that $P>0$ when pushing the right colloid towards the left one).
To compute the pressure, we can then use the contact theorem \eqref{CT_I}
at $x=0$ (left colloid position). The attractive field term is $-1/4$ and the repulsive kinetic term is given by the right colloid's contact density. This density is exponential since the right colloid is subject to a constant force $-1/4 -\widetilde P$. Therefore, the suitably normalized right colloid density profile reads:
\begin{equation}
    \widetilde{n}(\widetilde{x}_{C_R}) = 
    \left(\widetilde P + \frac{1}{4} \right)
    \e^{-(\widetilde P+1/4)\widetilde x_{\!_R}}
\end{equation}
where $\widetilde{x}_{C_R} \geq  0$ is the position of the right colloid and $\braket{\widetilde{L}}$ its average position. The contact theorem then holds trivially:
\begin{equation}
\widetilde{P} = \widetilde n(0) -\frac{1}{4}
\label{eq:app_eos2}
\end{equation}
and a by-product of the argument is that 
\begin{equation}
    \left\langle\widetilde x_{\!R} \right\rangle \,=\, \braket{\widetilde{L}} \, = 
     \left(\widetilde P + \frac{1}{4} \right)^{-1}.
\end{equation}
The isobaric equation of state \eqref{eq:app_eos1} is thereby recovered.
The similar form taken by the canonical pressure for ($N=1$, $\Delta=1$) and isobaric pressure for ($N=1$, $\Delta=-1$)
is therefore coincidental: the kinetic contributions to
the pressure stem from different ionic and colloid profiles. 

\section{Family of constants}
\label{appx:familyofconstants}
	 
The following coefficients hold for all $N$:
\begin{align}
a_{nmk} &= \frac{2^{1-\delta _{nm}} \left(\frac{1-\Delta }{1+\Delta}\right)^{m+n}\left(i\left(\frac{N}{2}-k\right)\right)^{-2\delta_{\frac{N}{2}\left\lfloor \frac{N}{2}\right\rfloor }}}{ (m!n!)^2 (m-k)_{n-m} (k+m-N)_{n-m}}  \nonumber\\ 
\;\times &\frac{(2k-N)^2}{(k+n-N)^2_{\left\lceil \frac{N}{2}\right\rceil -n}(n-k)_{k-n}^2 \Gamma \left(\left\lceil \frac{N}{2}\right\rceil -k\right)^2},
\end{align}

\begin{align}
&b_{nmk}=\frac{2a_{nmk}}{N-2k} \left(\frac{H_{N-k-m}-H_{k-m}+H_{N-k-n}-H_{k-n}}{2}\nonumber\right.\\ 
&\left. + H_{\ceil{\frac{N}{2}}-k -1}- H_{\floor{\frac{N}{2}}-k}
+\delta _{kn}H_{k-n}-\frac{1-\delta _{\frac{N}{2}\left\lfloor \frac{N}{2}\right\rfloor }}{\frac{N}{2}-k}\right),
\end{align}

\begin{align}
c_{nmk}=\frac{2^{1-\delta _{nm}} (m!n!)^{-2}(k+m-N)_{k-m}^{-1}\left(\frac{1-\Delta }{1+\Delta}\right)^{m+n}}{ (2 k-N+1)_{N-k-n}(m-k)_{k-m} (N-k-n)!},
\end{align}
where $H_n$ is the n-th harmonic number, $(x)_n = \Gamma(x+n)/\Gamma(x)$ is the Pochhammer symbol and $(x)_n^2 =((x)_n)^2$. The Kronecker delta $\delta _{\frac{N}{2}\left\lfloor \frac{N}{2}\right\rfloor }$ is used for terms that appear exclusively when $N$ is even.

\section{Contact theorem: two derivations}
\label{sec:CT}
			
The structure of the partition function and the density profile is given by the sum of all possible values of $({N_{\ell}, N_r})$, such that $N_{\ell} + N_r \leq N$, each with its respective weight given by $z_{N_{\ell},N_r}(N,\Delta,\widetilde{L}) /Z(N,\Delta,\widetilde{L})$. In a similar fashion as in sections \ref{subsec:1cp} and \ref{subsec:Ncp}, the contact condition for permeable colloids follows from the impermeable cases. From previous works (see \cite{vtt2016}), we already know the contact theorem for impermeable colloids:
\begin{equation}
	\widetilde{P}_{N_r, N_\ell} = \widetilde{n}(0^+) - (N/2-N_{\ell})^2,
	\label{CT_I}		
\end{equation}
where $\widetilde{P}_{N_r, N_\ell}$ is the canonical pressure for a system with $N_{\ell}$ and $N_r$ counter-ions in the left and right regions respectively. In this section, the indexes regarding the type of ensemble are omitted since we will discuss exclusively the canonical functions. By averaging equation \eqref{CT_I} over all possible configurations of $N$ counterions with a fixed number of counterions in the left and right regions, we get a contact condition:
\begin{equation}
	\widetilde{P} = \widetilde{n}(0) - \left<(N/2-N_{\ell})^2 \right>,
	\label{CT_P}
\end{equation}
where we have used the fact that density is continuous for the permeable case and thus $\widetilde{n}(0^{\pm}) = \widetilde{n}(0)$. This expression can be stated in terms of average number of counterions by using the contact theorem for the exterior region in the impermeable case  $\widetilde{n}(0^-) =  (\epsilon_2/\epsilon_1)N_{\ell}^2$. Upon averaging we get
\begin{equation}
	\widetilde{n}(0) =\frac{1+\Delta}{1-\Delta}\left<N_{\ell}^2 \right>,
	\label{CT_outsidedensity}
\end{equation}
where $\widetilde{n}(0^{-}) = \widetilde{n}(0)$ was used again. The previous result allows to express the pressure obtained in equation \eqref{CT_P} in terms of the moments of the total charge in the left region: 
\begin{equation}
	\widetilde{P} = \frac{1+\Delta}{1- \Delta}\left<N_{\ell}^2 \right> - \left<\left(\frac{N}{2}-N_{\ell}\right)^2 \right>,
	\label{CT_DP}
\end{equation}
which if expanded yields equation \eqref{ConTeoNr}. 

Let us now turn to a derivation of a contact relation from a mechanical approach. We start from the stress tensor  component $T_{{x}{x}}$ which in 1D is  given by \cite{jackson}:
\begin{equation}
	{T}_{{x}{x}} =- k T{n}({x}) + \frac{\epsilon({x})}{4} {E}^2({x}).
\end{equation} 
We define the dimensionless stress tensor as $\widetilde{T}_{\widetilde{x}\widetilde{x}} = {T}_{xx} \epsilon_2/e^2$, which in turn will give us a dimensionless force (pressure in 1D) $\widetilde{F} = \epsilon_2{F}/e^2$:
\begin{equation}
	\widetilde{T}_{\widetilde{x}\widetilde{x}} = -\widetilde{n}(\widetilde{x}) + \frac{\epsilon(\widetilde{x})}{4\epsilon_2} \widetilde{E}^2(\widetilde{x}),
\end{equation} 
where $\widetilde{E} =  \epsilon_2 E/e$. The force on the colloid is given by $\widetilde{F} = \widetilde{T}_{\widetilde{x}\widetilde{x}}(0^+) - \widetilde{T}_{\widetilde{x}\widetilde{x}}(0^-)$. Then we take the average which gives:
\begin{equation}
	\braket{\widetilde{F}} = \left< \widetilde{n}(0^-) - \widetilde{n}(0^+) -  \frac{1}{4}\left(\frac{\epsilon_1}{\epsilon_2} \widetilde{E}^2(0^-) -  \widetilde{E}^2(0^+)\right)\right>.
\end{equation}	
The permeability condition yields $\left< \widetilde{n}(0^-) \right> =\left< \widetilde{n}(0^+)\right>$,
leaving only the electric contributions:
\begin{equation}
	\braket{\widetilde{F}} = \frac{1}{4}\left<\widetilde{E}^2(0^+) -\frac{1-\Delta}{1+\Delta} \widetilde{E}^2(0^-)\right>.
	\label{CT_ForceD}
\end{equation}
The fields are $\widetilde{E}(0^-) = 2N_{\ell} (1+\Delta)/(1-\Delta)$ and $\widetilde{E}(0^+) = (2N_{\ell}-N)$ which leads to:
\begin{align}
	\braket{\widetilde{F}} = -\frac{1+\Delta}{1- \Delta}\left<N_{\ell}^2 \right> + \left<\left(\frac{N}{2}-N_{\ell}\right)^2 \right>,
	\label{CT9}
\end{align}
and \eqref{CT_DP} is recovered, but for the sign convention (a positive pressure corresponds to a negative force acting on the leftmost charge). Note that $\Delta = 0$ yields a special case in which the second moment does not contribute to the pressure:
\begin{equation}
	\braket{\widetilde{F}}(\Delta = 0) =   \frac{N^2}{4} - N\braket{N_{\ell}}  .
\end{equation}

We conclude with a note on the similarity in terms of functional form of the contact relation presented here, with one of the few known formulas for a contact-like relation in the presence of a dielectric discontinuity. Carnie and Chan (\cite{carnie_chan}, Eq. 3.26) derived such an equation for an electrolyte with planar geometry. Unfortunately, these results are for a single impenetrable wall/colloid with a dielectric jump at the interface, for which a counter-ion solution yields a vanishing pressure. Remarkably, Eq. 3.26 has the same functional form as $P_{N_r, N_\ell}$, predicted by equation \eqref{CT_I}. 
The difference lies in the density profile formed around the colloid or more precisely at contact, $n(0^+)$: for the single wall $n(0^+) = \epsilon_2 E(0^+)^2/8\pi k T$  unlike the contact density discussed in section \ref{sec:Nimpermeable}, which has a dependence on $N_{\ell}, N_{r}$ and $L$
(explicit expressions found in \cite{vtt2016}). Besides, another difference between our results and those reported in
\cite{carnie_chan} is in the relevance of the parity of $N$ (say in the impermeable situation, with all counterions in the middle region and $L\to \infty$, that is the closest to the infinite length 
geometry worked out in \cite{carnie_chan}):
this is a peculiarity of the one-dimensional setting, where the colloids may attract
with a finite force even when $L\to\infty$.


\bibliography{refs}

\begin{thebibliography}{48}%
\makeatletter
\providecommand \@ifxundefined [1]{%
 \@ifx{#1\undefined}
}%
\providecommand \@ifnum [1]{%
 \ifnum #1\expandafter \@firstoftwo
 \else \expandafter \@secondoftwo
 \fi
}%
\providecommand \@ifx [1]{%
 \ifx #1\expandafter \@firstoftwo
 \else \expandafter \@secondoftwo
 \fi
}%
\providecommand \natexlab [1]{#1}%
\providecommand \enquote  [1]{``#1''}%
\providecommand \bibnamefont  [1]{#1}%
\providecommand \bibfnamefont [1]{#1}%
\providecommand \citenamefont [1]{#1}%
\providecommand \href@noop [0]{\@secondoftwo}%
\providecommand \href [0]{\begingroup \@sanitize@url \@href}%
\providecommand \@href[1]{\@@startlink{#1}\@@href}%
\providecommand \@@href[1]{\endgroup#1\@@endlink}%
\providecommand \@sanitize@url [0]{\catcode `\\12\catcode `\$12\catcode
  `\&12\catcode `\#12\catcode `\^12\catcode `\_12\catcode `\%12\relax}%
\providecommand \@@startlink[1]{}%
\providecommand \@@endlink[0]{}%
\providecommand \url  [0]{\begingroup\@sanitize@url \@url }%
\providecommand \@url [1]{\endgroup\@href {#1}{\urlprefix }}%
\providecommand \urlprefix  [0]{URL }%
\providecommand \Eprint [0]{\href }%
\providecommand \doibase [0]{https://doi.org/}%
\providecommand \selectlanguage [0]{\@gobble}%
\providecommand \bibinfo  [0]{\@secondoftwo}%
\providecommand \bibfield  [0]{\@secondoftwo}%
\providecommand \translation [1]{[#1]}%
\providecommand \BibitemOpen [0]{}%
\providecommand \bibitemStop [0]{}%
\providecommand \bibitemNoStop [0]{.\EOS\space}%
\providecommand \EOS [0]{\spacefactor3000\relax}%
\providecommand \BibitemShut  [1]{\csname bibitem#1\endcsname}%
\let\auto@bib@innerbib\@empty
\bibitem [{\citenamefont {Holm}\ \emph {et~al.}(2001)\citenamefont {Holm},
  \citenamefont {K{\'e}kicheff},\ and\ \citenamefont {Podgornik}}]{Holm2001}%
  \BibitemOpen
  \bibfield  {author} {\bibinfo {author} {\bibfnamefont {C.}~\bibnamefont
  {Holm}}, \bibinfo {author} {\bibfnamefont {P.}~\bibnamefont
  {K{\'e}kicheff}},\ and\ \bibinfo {author} {\bibfnamefont {R.}~\bibnamefont
  {Podgornik}},\ }\href
  {https://doi.org/https://doi.org/10.1007/978-94-010-0577-7} {\emph {\bibinfo
  {title} {Electrostatic Effects in Soft Matter and Biophysics}}}\ (\bibinfo
  {publisher} {Kluwer Academic, Dordrecht},\ \bibinfo {year}
  {2001})\BibitemShut {NoStop}%
\bibitem [{\citenamefont {Andelman}(2006)}]{andelman2006}%
  \BibitemOpen
  \bibfield  {author} {\bibinfo {author} {\bibfnamefont {D.}~\bibnamefont
  {Andelman}},\ }\bibinfo {title} {Introduction to electrostatics in soft and
  biological matter},\ in\ \href
  {https://books.google.com.co/books?id=hk9UbvyKZE0C} {\emph {\bibinfo
  {booktitle} {Soft Condensed Matter Physics in Molecular and Cell Biology}}},\
  \bibinfo {editor} {edited by\ \bibinfo {editor} {\bibfnamefont {W.~C.~K.}\
  \bibnamefont {Poon}}\ and\ \bibinfo {editor} {\bibfnamefont {D.}~\bibnamefont
  {Andelman}}}\ (\bibinfo  {publisher} {Taylor \& Francis, New York},\ \bibinfo
  {year} {2006})\ pp.\ \bibinfo {pages} {97--122}\BibitemShut {NoStop}%
\bibitem [{\citenamefont {Levin}(2002)}]{Levin2002}%
  \BibitemOpen
  \bibfield  {author} {\bibinfo {author} {\bibfnamefont {Y.}~\bibnamefont
  {Levin}},\ }\bibfield  {title} {\bibinfo {title} {Electrostatic correlations:
  from plasma to biology},\ }\href
  {https://doi.org/10.1088/0034-4885/65/11/201} {\bibfield  {journal} {\bibinfo
   {journal} {Rep. Pro. Phys.}\ }\textbf {\bibinfo {volume} {65}},\ \bibinfo
  {pages} {1577} (\bibinfo {year} {2002})}\BibitemShut {NoStop}%
\bibitem [{\citenamefont {Naji}\ \emph {et~al.}(2005)\citenamefont {Naji},
  \citenamefont {Jungblut}, \citenamefont {Moreira},\ and\ \citenamefont
  {Netz}}]{Naji2005}%
  \BibitemOpen
  \bibfield  {author} {\bibinfo {author} {\bibfnamefont {A.}~\bibnamefont
  {Naji}}, \bibinfo {author} {\bibfnamefont {S.}~\bibnamefont {Jungblut}},
  \bibinfo {author} {\bibfnamefont {A.~G.}\ \bibnamefont {Moreira}},\ and\
  \bibinfo {author} {\bibfnamefont {R.~R.}\ \bibnamefont {Netz}},\ }\bibfield
  {title} {\bibinfo {title} {Electrostatic interactions in strongly coupled
  soft matter},\ }\href
  {https://doi.org/https://doi.org/10.1016/j.physa.2004.12.029} {\bibfield
  {journal} {\bibinfo  {journal} {Physica A}\ }\textbf {\bibinfo {volume}
  {352}},\ \bibinfo {pages} {131 } (\bibinfo {year} {2005})}\BibitemShut
  {NoStop}%
\bibitem [{\citenamefont {Boroudjerdi}\ \emph {et~al.}(2005)\citenamefont
  {Boroudjerdi}, \citenamefont {Kim}, \citenamefont {Naji}, \citenamefont
  {Netz}, \citenamefont {Schlagberger},\ and\ \citenamefont
  {Serr}}]{Boroudjerdi2005}%
  \BibitemOpen
  \bibfield  {author} {\bibinfo {author} {\bibfnamefont {H.}~\bibnamefont
  {Boroudjerdi}}, \bibinfo {author} {\bibfnamefont {Y.-W.}\ \bibnamefont
  {Kim}}, \bibinfo {author} {\bibfnamefont {A.}~\bibnamefont {Naji}}, \bibinfo
  {author} {\bibfnamefont {R.~R.}\ \bibnamefont {Netz}}, \bibinfo {author}
  {\bibfnamefont {X.}~\bibnamefont {Schlagberger}},\ and\ \bibinfo {author}
  {\bibfnamefont {A.}~\bibnamefont {Serr}},\ }\bibfield  {title} {\bibinfo
  {title} {Statics and dynamics of strongly charged soft matter},\ }\href
  {https://doi.org/https://doi.org/10.1016/j.physrep.2005.06.006} {\bibfield
  {journal} {\bibinfo  {journal} {Phys. Rep.}\ }\textbf {\bibinfo {volume}
  {416}},\ \bibinfo {pages} {129 } (\bibinfo {year} {2005})}\BibitemShut
  {NoStop}%
\bibitem [{\citenamefont {Ioannidou}\ \emph {et~al.}(2016)\citenamefont
  {Ioannidou}, \citenamefont {Kandu{\v{c}}}, \citenamefont {Li}, \citenamefont
  {Frenkel}, \citenamefont {Dobnikar},\ and\ \citenamefont
  {Del~Gado}}]{Ioannidou2016}%
  \BibitemOpen
  \bibfield  {author} {\bibinfo {author} {\bibfnamefont {K.}~\bibnamefont
  {Ioannidou}}, \bibinfo {author} {\bibfnamefont {M.}~\bibnamefont
  {Kandu{\v{c}}}}, \bibinfo {author} {\bibfnamefont {L.}~\bibnamefont {Li}},
  \bibinfo {author} {\bibfnamefont {D.}~\bibnamefont {Frenkel}}, \bibinfo
  {author} {\bibfnamefont {J.}~\bibnamefont {Dobnikar}},\ and\ \bibinfo
  {author} {\bibfnamefont {E.}~\bibnamefont {Del~Gado}},\ }\bibfield  {title}
  {\bibinfo {title} {The crucial effect of early-stage gelation on the
  mechanical properties of cement hydrates},\ }\href
  {https://doi.org/10.1038/ncomms12106} {\bibfield  {journal} {\bibinfo
  {journal} {Nat. Commun.}\ }\textbf {\bibinfo {volume} {7}},\ \bibinfo {pages}
  {12106} (\bibinfo {year} {2016})}\BibitemShut {NoStop}%
\bibitem [{\citenamefont {Harris}(1836)}]{Harris}%
  \BibitemOpen
  \bibfield  {author} {\bibinfo {author} {\bibfnamefont {W.~S.}\ \bibnamefont
  {Harris}},\ }\bibfield  {title} {\bibinfo {title} {Xx. inquiries concerning
  the elementary laws of electricity. second series},\ }\href
  {https://doi.org/https://doi.org/10.1098/rstl.1836.0022} {\bibfield
  {journal} {\bibinfo  {journal} {Trans. R. Soc. London}\ }\textbf {\bibinfo
  {volume} {126}},\ \bibinfo {pages} {417} (\bibinfo {year}
  {1836})}\BibitemShut {NoStop}%
\bibitem [{\citenamefont {Lekner}(2012)}]{Lekner}%
  \BibitemOpen
  \bibfield  {author} {\bibinfo {author} {\bibfnamefont {J.}~\bibnamefont
  {Lekner}},\ }\bibfield  {title} {\bibinfo {title} {Electrostatics of two
  charged conducting spheres},\ }\href
  {https://doi.org/https://doi.org/10.1098/rspa.2012.0133} {\bibfield
  {journal} {\bibinfo  {journal} {Proc. R. Soc. A}\ }\textbf {\bibinfo {volume}
  {468}},\ \bibinfo {pages} {2829} (\bibinfo {year} {2012})}\BibitemShut
  {NoStop}%
\bibitem [{\citenamefont {dos Santos}\ and\ \citenamefont
  {Levin}(2019)}]{Alexandre19}%
  \BibitemOpen
  \bibfield  {author} {\bibinfo {author} {\bibfnamefont {A.~P.}\ \bibnamefont
  {dos Santos}}\ and\ \bibinfo {author} {\bibfnamefont {Y.}~\bibnamefont
  {Levin}},\ }\bibfield  {title} {\bibinfo {title} {Like-charge attraction
  between metal nanoparticles in a $1\ensuremath{\mathbin:}1$ electrolyte
  solution},\ }\href
  {https://doi.org/https://doi.org/10.1103/PhysRevLett.122.248005} {\bibfield
  {journal} {\bibinfo  {journal} {Phys. Rev. Lett.}\ }\textbf {\bibinfo
  {volume} {122}},\ \bibinfo {pages} {248005} (\bibinfo {year}
  {2019})}\BibitemShut {NoStop}%
\bibitem [{\citenamefont {{R. R. Netz}}(2001)}]{netzSCtoPB}%
  \BibitemOpen
  \bibfield  {author} {\bibinfo {author} {\bibnamefont {{R. R. Netz}}},\
  }\bibfield  {title} {\bibinfo {title} {Electrostatistics of counter-ions at
  and between planar charged walls: From poisson-boltzmann to the
  strong-coupling theory},\ }\href {https://doi.org/epje/v5/p557(e01021)}
  {\bibfield  {journal} {\bibinfo  {journal} {Eur. Phys. J. E}\ }\textbf
  {\bibinfo {volume} {5}},\ \bibinfo {pages} {557} (\bibinfo {year}
  {2001})}\BibitemShut {NoStop}%
\bibitem [{\citenamefont {{\v{S}}amaj}\ and\ \citenamefont
  {Trizac}(2011)}]{SaTr11PRL}%
  \BibitemOpen
  \bibfield  {author} {\bibinfo {author} {\bibfnamefont {L.}~\bibnamefont
  {{\v{S}}amaj}}\ and\ \bibinfo {author} {\bibfnamefont {E.}~\bibnamefont
  {Trizac}},\ }\bibfield  {title} {\bibinfo {title} {Counterions at highly
  charged interfaces: From one plate to like-charge attraction},\ }\href
  {https://doi.org/https://doi.org/10.1103/PhysRevLett.106.078301} {\bibfield
  {journal} {\bibinfo  {journal} {Phys. Rev. Lett.}\ }\textbf {\bibinfo
  {volume} {106}},\ \bibinfo {pages} {078301} (\bibinfo {year}
  {2011})}\BibitemShut {NoStop}%
\bibitem [{\citenamefont {{\v{S}}amaj}\ \emph {et~al.}(2018)\citenamefont
  {{\v{S}}amaj}, \citenamefont {Trulsson},\ and\ \citenamefont
  {Trizac}}]{SaTT18}%
  \BibitemOpen
  \bibfield  {author} {\bibinfo {author} {\bibfnamefont {L.}~\bibnamefont
  {{\v{S}}amaj}}, \bibinfo {author} {\bibfnamefont {M.}~\bibnamefont
  {Trulsson}},\ and\ \bibinfo {author} {\bibfnamefont {E.}~\bibnamefont
  {Trizac}},\ }\bibfield  {title} {\bibinfo {title} {Strong-coupling theory of
  counterions between symmetrically charged walls: From crystal to fluid
  phases},\ }\href {https://doi.org/https://doi.org/10.1039/c8sm00571k}
  {\bibfield  {journal} {\bibinfo  {journal} {Soft Matter}\ }\textbf {\bibinfo
  {volume} {14}},\ \bibinfo {pages} {4040} (\bibinfo {year}
  {2018})}\BibitemShut {NoStop}%
\bibitem [{\citenamefont {Neu}(1999)}]{Neu1999}%
  \BibitemOpen
  \bibfield  {author} {\bibinfo {author} {\bibfnamefont {J.~C.}\ \bibnamefont
  {Neu}},\ }\bibfield  {title} {\bibinfo {title} {Wall-mediated forces between
  like-charged bodies in an electrolyte},\ }\href
  {https://doi.org/https://doi.org/10.1103/PhysRevLett.82.1072} {\bibfield
  {journal} {\bibinfo  {journal} {Phys. Rev. Lett.}\ }\textbf {\bibinfo
  {volume} {82}},\ \bibinfo {pages} {1072} (\bibinfo {year}
  {1999})}\BibitemShut {NoStop}%
\bibitem [{\citenamefont {K\'ekicheff}\ \emph {et~al.}(1993)\citenamefont
  {K\'ekicheff}, \citenamefont {Mar{\v{c}}elja}, \citenamefont {Senden},\ and\
  \citenamefont {Shubin}}]{Kekicheff1993}%
  \BibitemOpen
  \bibfield  {author} {\bibinfo {author} {\bibfnamefont {P.}~\bibnamefont
  {K\'ekicheff}}, \bibinfo {author} {\bibfnamefont {S.}~\bibnamefont
  {Mar{\v{c}}elja}}, \bibinfo {author} {\bibfnamefont {T.~J.}\ \bibnamefont
  {Senden}},\ and\ \bibinfo {author} {\bibfnamefont {V.~E.}\ \bibnamefont
  {Shubin}},\ }\bibfield  {title} {\bibinfo {title} {Charge reversal seen in
  electrical double layer interaction of surfaces immersed in 2:1 calcium
  electrolyte},\ }\href {https://doi.org/https://doi.org/10.1063/1.465906}
  {\bibfield  {journal} {\bibinfo  {journal} {J. Chem. Phys.}\ }\textbf
  {\bibinfo {volume} {99}},\ \bibinfo {pages} {6098} (\bibinfo {year}
  {1993})}\BibitemShut {NoStop}%
\bibitem [{\citenamefont {Crocker}\ and\ \citenamefont
  {Grier}(1996)}]{crockerGrier}%
  \BibitemOpen
  \bibfield  {author} {\bibinfo {author} {\bibfnamefont {J.~C.}\ \bibnamefont
  {Crocker}}\ and\ \bibinfo {author} {\bibfnamefont {D.~G.}\ \bibnamefont
  {Grier}},\ }\bibfield  {title} {\bibinfo {title} {When like charges attract:
  The effects of geometrical confinement on long-range colloidal
  interactions},\ }\href
  {https://doi.org/https://doi.org/10.1103/PhysRevLett.77.1897} {\bibfield
  {journal} {\bibinfo  {journal} {Phys. Rev. Lett.}\ }\textbf {\bibinfo
  {volume} {77}},\ \bibinfo {pages} {1897} (\bibinfo {year}
  {1996})}\BibitemShut {NoStop}%
\bibitem [{\citenamefont {Kepler}\ and\ \citenamefont
  {Fraden}(1994)}]{keplerFraden}%
  \BibitemOpen
  \bibfield  {author} {\bibinfo {author} {\bibfnamefont {G.~M.}\ \bibnamefont
  {Kepler}}\ and\ \bibinfo {author} {\bibfnamefont {S.}~\bibnamefont
  {Fraden}},\ }\bibfield  {title} {\bibinfo {title} {Attractive potential
  between confined colloids at low ionic strength},\ }\href
  {https://doi.org/https://doi.org/10.1103/PhysRevLett.73.356} {\bibfield
  {journal} {\bibinfo  {journal} {Phys. Rev. Lett.}\ }\textbf {\bibinfo
  {volume} {73}},\ \bibinfo {pages} {356} (\bibinfo {year} {1994})}\BibitemShut
  {NoStop}%
\bibitem [{\citenamefont {Allahyarov}\ \emph {et~al.}(1999)\citenamefont
  {Allahyarov}, \citenamefont {D'Amico},\ and\ \citenamefont
  {L\"owen}}]{AllahyarovDamicoLowen}%
  \BibitemOpen
  \bibfield  {author} {\bibinfo {author} {\bibfnamefont {E.}~\bibnamefont
  {Allahyarov}}, \bibinfo {author} {\bibfnamefont {I.}~\bibnamefont
  {D'Amico}},\ and\ \bibinfo {author} {\bibfnamefont {H.}~\bibnamefont
  {L\"owen}},\ }\bibfield  {title} {\bibinfo {title} {Effect of geometrical
  confinement on the interaction between charged colloidal suspensions},\
  }\href {https://doi.org/https://doi.org/10.1103/PhysRevE.60.3199} {\bibfield
  {journal} {\bibinfo  {journal} {Phys. Rev. E}\ }\textbf {\bibinfo {volume}
  {60}},\ \bibinfo {pages} {3199} (\bibinfo {year} {1999})}\BibitemShut
  {NoStop}%
\bibitem [{\citenamefont {Grønbech-Jensen}\ \emph {et~al.}(1998)\citenamefont
  {Grønbech-Jensen}, \citenamefont {Beardmore},\ and\ \citenamefont
  {Pincus}}]{GRONBECHJENSEN}%
  \BibitemOpen
  \bibfield  {author} {\bibinfo {author} {\bibfnamefont {N.}~\bibnamefont
  {Grønbech-Jensen}}, \bibinfo {author} {\bibfnamefont {K.~M.}\ \bibnamefont
  {Beardmore}},\ and\ \bibinfo {author} {\bibfnamefont {P.}~\bibnamefont
  {Pincus}},\ }\bibfield  {title} {\bibinfo {title} {Interactions between
  charged spheres in divalent counterion solution},\ }\href
  {https://doi.org/https://doi.org/10.1016/S0378-4371(98)00369-0} {\bibfield
  {journal} {\bibinfo  {journal} {Physica A}\ }\textbf {\bibinfo {volume}
  {261}},\ \bibinfo {pages} {74 } (\bibinfo {year} {1998})}\BibitemShut
  {NoStop}%
\bibitem [{\citenamefont {Ma}\ \emph {et~al.}(2001)\citenamefont {Ma},
  \citenamefont {Girvin},\ and\ \citenamefont
  {Rajaraman}}]{ningGirvinRajaraman}%
  \BibitemOpen
  \bibfield  {author} {\bibinfo {author} {\bibfnamefont {N.}~\bibnamefont
  {Ma}}, \bibinfo {author} {\bibfnamefont {S.~M.}\ \bibnamefont {Girvin}},\
  and\ \bibinfo {author} {\bibfnamefont {R.}~\bibnamefont {Rajaraman}},\
  }\bibfield  {title} {\bibinfo {title} {Effective attraction between
  like-charged colloids in a two-dimensional plasma},\ }\href
  {https://doi.org/https://doi.org/10.1103/PhysRevE.63.021402} {\bibfield
  {journal} {\bibinfo  {journal} {Phys. Rev. E}\ }\textbf {\bibinfo {volume}
  {63}},\ \bibinfo {pages} {021402} (\bibinfo {year} {2001})}\BibitemShut
  {NoStop}%
\bibitem [{\citenamefont {Guldbrand}\ \emph {et~al.}(1984)\citenamefont
  {Guldbrand}, \citenamefont {Jönsson}, \citenamefont {Wennerström},\ and\
  \citenamefont {Linse}}]{Guldbrand1984}%
  \BibitemOpen
  \bibfield  {author} {\bibinfo {author} {\bibfnamefont {L.}~\bibnamefont
  {Guldbrand}}, \bibinfo {author} {\bibfnamefont {B.}~\bibnamefont {Jönsson}},
  \bibinfo {author} {\bibfnamefont {H.}~\bibnamefont {Wennerström}},\ and\
  \bibinfo {author} {\bibfnamefont {P.}~\bibnamefont {Linse}},\ }\bibfield
  {title} {\bibinfo {title} {Electrical double layer forces. a {Monte Carlo}
  study},\ }\href {https://doi.org/https://doi.org/10.1063/1.446912} {\bibfield
   {journal} {\bibinfo  {journal} {J. Chem. Phys.}\ }\textbf {\bibinfo {volume}
  {80}},\ \bibinfo {pages} {2221} (\bibinfo {year} {1984})}\BibitemShut
  {NoStop}%
\bibitem [{\citenamefont {Moreira}\ and\ \citenamefont
  {Netz}(2002)}]{Moreira2002MC}%
  \BibitemOpen
  \bibfield  {author} {\bibinfo {author} {\bibfnamefont {A.~G.}\ \bibnamefont
  {Moreira}}\ and\ \bibinfo {author} {\bibfnamefont {R.~R.}\ \bibnamefont
  {Netz}},\ }\bibfield  {title} {\bibinfo {title} {Simulations of counterions
  at charged plates},\ }\href
  {https://doi.org/https://doi.org/10.1140/epje/i2001-10091-9} {\bibfield
  {journal} {\bibinfo  {journal} {Euro. Phys. J. E}\ }\textbf {\bibinfo
  {volume} {8}},\ \bibinfo {pages} {33} (\bibinfo {year} {2002})}\BibitemShut
  {NoStop}%
\bibitem [{\citenamefont {Jho}\ \emph {et~al.}(2008)\citenamefont {Jho},
  \citenamefont {Kandu{\v{c}}}, \citenamefont {Naji}, \citenamefont
  {Podgornik}, \citenamefont {Kim},\ and\ \citenamefont {Pincus}}]{Jho2008}%
  \BibitemOpen
  \bibfield  {author} {\bibinfo {author} {\bibfnamefont {Y.~S.}\ \bibnamefont
  {Jho}}, \bibinfo {author} {\bibfnamefont {M.}~\bibnamefont {Kandu{\v{c}}}},
  \bibinfo {author} {\bibfnamefont {A.}~\bibnamefont {Naji}}, \bibinfo {author}
  {\bibfnamefont {R.}~\bibnamefont {Podgornik}}, \bibinfo {author}
  {\bibfnamefont {M.~W.}\ \bibnamefont {Kim}},\ and\ \bibinfo {author}
  {\bibfnamefont {P.~A.}\ \bibnamefont {Pincus}},\ }\bibfield  {title}
  {\bibinfo {title} {Strong-coupling electrostatics in the presence of
  dielectric inhomogeneities},\ }\href
  {https://doi.org/https://doi.org/10.1103/physrevlett.101.188101} {\bibfield
  {journal} {\bibinfo  {journal} {Phys. Rev. Lett.}\ }\textbf {\bibinfo
  {volume} {101}},\ \bibinfo {pages} {188101} (\bibinfo {year}
  {2008})}\BibitemShut {NoStop}%
\bibitem [{\citenamefont {Kjellander}\ and\ \citenamefont
  {Mar\v{c}elja}(1984)}]{Kjellander1984}%
  \BibitemOpen
  \bibfield  {author} {\bibinfo {author} {\bibfnamefont {R.}~\bibnamefont
  {Kjellander}}\ and\ \bibinfo {author} {\bibfnamefont {S.}~\bibnamefont
  {Mar\v{c}elja}},\ }\bibfield  {title} {\bibinfo {title} {Correlation and
  image charge effects in electric double layers},\ }\href
  {https://doi.org/https://doi.org/10.1016/0009-2614(84)87039-6} {\bibfield
  {journal} {\bibinfo  {journal} {Chem. Phys. Lett.}\ }\textbf {\bibinfo
  {volume} {112}},\ \bibinfo {pages} {49 } (\bibinfo {year}
  {1984})}\BibitemShut {NoStop}%
\bibitem [{\citenamefont {Kandu{\v{c}}}\ and\ \citenamefont
  {Podgornik}(2007)}]{Kanduc2007}%
  \BibitemOpen
  \bibfield  {author} {\bibinfo {author} {\bibfnamefont {M.}~\bibnamefont
  {Kandu{\v{c}}}}\ and\ \bibinfo {author} {\bibfnamefont {R.}~\bibnamefont
  {Podgornik}},\ }\bibfield  {title} {\bibinfo {title} {Electrostatic image
  effects for counterions between charged planar walls},\ }\href
  {https://doi.org/https://doi.org/10.1140/epje/i2007-10187-2} {\bibfield
  {journal} {\bibinfo  {journal} {Eur. Phys. J. E, Soft matter}\ }\textbf
  {\bibinfo {volume} {23}},\ \bibinfo {pages} {265—274} (\bibinfo {year}
  {2007})}\BibitemShut {NoStop}%
\bibitem [{\citenamefont {{\v{S}}amaj}\ and\ \citenamefont
  {Trizac}(2012{\natexlab{a}})}]{SaTr2012EPL}%
  \BibitemOpen
  \bibfield  {author} {\bibinfo {author} {\bibfnamefont {L.}~\bibnamefont
  {{\v{S}}amaj}}\ and\ \bibinfo {author} {\bibfnamefont {E.}~\bibnamefont
  {Trizac}},\ }\bibfield  {title} {\bibinfo {title} {Ground state of classical
  bilayer {Wigner} crystals},\ }\href
  {https://doi.org/10.1209/0295-5075/98/36004} {\bibfield  {journal} {\bibinfo
  {journal} {Europhys. Lett.}\ }\textbf {\bibinfo {volume} {98}},\ \bibinfo
  {pages} {36004} (\bibinfo {year} {2012}{\natexlab{a}})}\BibitemShut {NoStop}%
\bibitem [{\citenamefont {{\v{S}}amaj}\ and\ \citenamefont
  {Trizac}(2012{\natexlab{b}})}]{SaTr2012CPP}%
  \BibitemOpen
  \bibfield  {author} {\bibinfo {author} {\bibfnamefont {L.}~\bibnamefont
  {{\v{S}}amaj}}\ and\ \bibinfo {author} {\bibfnamefont {E.}~\bibnamefont
  {Trizac}},\ }\bibfield  {title} {\bibinfo {title} {Strong‐coupling theory
  for a polarizable planar colloid},\ }\href
  {https://doi.org/https://doi.org/10.1002/ctpp.201100059} {\bibfield
  {journal} {\bibinfo  {journal} {Contrib. Plasma Phys.}\ }\textbf {\bibinfo
  {volume} {52}},\ \bibinfo {pages} {53} (\bibinfo {year}
  {2012}{\natexlab{b}})}\BibitemShut {NoStop}%
\bibitem [{\citenamefont {\v{S}amaj}\ \emph {et~al.}(2016)\citenamefont
  {\v{S}amaj}, \citenamefont {dos Santos}, \citenamefont {Levin},\ and\
  \citenamefont {Trizac}}]{Samaj2016}%
  \BibitemOpen
  \bibfield  {author} {\bibinfo {author} {\bibfnamefont {L.}~\bibnamefont
  {\v{S}amaj}}, \bibinfo {author} {\bibfnamefont {A.~P.}\ \bibnamefont {dos
  Santos}}, \bibinfo {author} {\bibfnamefont {Y.}~\bibnamefont {Levin}},\ and\
  \bibinfo {author} {\bibfnamefont {E.}~\bibnamefont {Trizac}},\ }\bibfield
  {title} {\bibinfo {title} {Mean-field beyond mean-field: the single particle
  view for moderately to strongly coupled charged fluids},\ }\href
  {https://doi.org/10.1039/C6SM01360K} {\bibfield  {journal} {\bibinfo
  {journal} {Soft Matter}\ }\textbf {\bibinfo {volume} {12}},\ \bibinfo {pages}
  {8768} (\bibinfo {year} {2016})}\BibitemShut {NoStop}%
\bibitem [{\citenamefont {Dean}\ \emph {et~al.}(1998)\citenamefont {Dean},
  \citenamefont {Horgan},\ and\ \citenamefont {Sentenac}}]{Dean1998}%
  \BibitemOpen
  \bibfield  {author} {\bibinfo {author} {\bibfnamefont {D.~S.}\ \bibnamefont
  {Dean}}, \bibinfo {author} {\bibfnamefont {R.~R.}\ \bibnamefont {Horgan}},\
  and\ \bibinfo {author} {\bibfnamefont {D.}~\bibnamefont {Sentenac}},\
  }\bibfield  {title} {\bibinfo {title} {Boundary effects in the
  one-dimensional {Coulomb} gas},\ }\href
  {https://doi.org/10.1023/A:1023241407140} {\bibfield  {journal} {\bibinfo
  {journal} {J. Stat. Phys.}\ }\textbf {\bibinfo {volume} {90}},\ \bibinfo
  {pages} {899} (\bibinfo {year} {1998})}\BibitemShut {NoStop}%
\bibitem [{\citenamefont {Dean}\ \emph {et~al.}(2009)\citenamefont {Dean},
  \citenamefont {Horgan},\ and\ \citenamefont {Podgornik}}]{Dean2009}%
  \BibitemOpen
  \bibfield  {author} {\bibinfo {author} {\bibfnamefont {D.~S.}\ \bibnamefont
  {Dean}}, \bibinfo {author} {\bibfnamefont {R.}~\bibnamefont {Horgan}},\ and\
  \bibinfo {author} {\bibfnamefont {R.}~\bibnamefont {Podgornik}},\ }\bibfield
  {title} {\bibinfo {title} {One-dimensional counterion gas between charged
  surfaces: Exact results compared with weak- and strong-coupling analyses},\
  }\href {https://doi.org/https://doi.org/10.1063/1.3078492} {\bibfield
  {journal} {\bibinfo  {journal} {J. Chem. Phys.}\ }\textbf {\bibinfo {volume}
  {130}},\ \bibinfo {pages} {094504} (\bibinfo {year} {2009})}\BibitemShut
  {NoStop}%
\bibitem [{\citenamefont {Démery}\ \emph {et~al.}(2012)\citenamefont
  {Démery}, \citenamefont {Dean}, \citenamefont {Hammant}, \citenamefont
  {Horgan},\ and\ \citenamefont {Podgornik}}]{demery2012}%
  \BibitemOpen
  \bibfield  {author} {\bibinfo {author} {\bibfnamefont {V.}~\bibnamefont
  {Démery}}, \bibinfo {author} {\bibfnamefont {D.~S.}\ \bibnamefont {Dean}},
  \bibinfo {author} {\bibfnamefont {T.~C.}\ \bibnamefont {Hammant}}, \bibinfo
  {author} {\bibfnamefont {R.~R.}\ \bibnamefont {Horgan}},\ and\ \bibinfo
  {author} {\bibfnamefont {R.}~\bibnamefont {Podgornik}},\ }\bibfield  {title}
  {\bibinfo {title} {The one-dimensional {Coulomb} lattice fluid capacitor},\
  }\href {https://doi.org/https://doi.org/10.1063/1.4740233} {\bibfield
  {journal} {\bibinfo  {journal} {J. Chem. Phys.}\ }\textbf {\bibinfo {volume}
  {137}},\ \bibinfo {pages} {064901} (\bibinfo {year} {2012})}\BibitemShut
  {NoStop}%
\bibitem [{\citenamefont {T\'ellez}\ and\ \citenamefont
  {Trizac}(2015)}]{tt2015}%
  \BibitemOpen
  \bibfield  {author} {\bibinfo {author} {\bibfnamefont {G.}~\bibnamefont
  {T\'ellez}}\ and\ \bibinfo {author} {\bibfnamefont {E.}~\bibnamefont
  {Trizac}},\ }\bibfield  {title} {\bibinfo {title} {Screening like charges in
  one-dimensional {Coulomb} systems: Exact results},\ }\href
  {https://doi.org/https://doi.org/10.1103/PhysRevE.92.042134} {\bibfield
  {journal} {\bibinfo  {journal} {Phys. Rev. E}\ }\textbf {\bibinfo {volume}
  {92}},\ \bibinfo {pages} {042134} (\bibinfo {year} {2015})}\BibitemShut
  {NoStop}%
\bibitem [{\citenamefont {Frydel}(2019)}]{frydel2019}%
  \BibitemOpen
  \bibfield  {author} {\bibinfo {author} {\bibfnamefont {D.}~\bibnamefont
  {Frydel}},\ }\bibfield  {title} {\bibinfo {title} {One-dimensional {Coulomb}
  system in a sticky wall confinement: Exact results},\ }\href
  {https://doi.org/https://doi.org/10.1103/PhysRevE.100.042113} {\bibfield
  {journal} {\bibinfo  {journal} {Phys. Rev. E}\ }\textbf {\bibinfo {volume}
  {100}},\ \bibinfo {pages} {042113} (\bibinfo {year} {2019})}\BibitemShut
  {NoStop}%
\bibitem [{\citenamefont {Lenard}(1961)}]{lenard}%
  \BibitemOpen
  \bibfield  {author} {\bibinfo {author} {\bibfnamefont {A.}~\bibnamefont
  {Lenard}},\ }\bibfield  {title} {\bibinfo {title} {Exact statistical
  mechanics of a one-dimensional system with {Coulomb} forces},\ }\href
  {https://doi.org/http://dx.doi.org/10.1063/1.1703757} {\bibfield  {journal}
  {\bibinfo  {journal} {J. Math. Phys.}\ }\textbf {\bibinfo {volume} {2}},\
  \bibinfo {pages} {682} (\bibinfo {year} {1961})}\BibitemShut {NoStop}%
\bibitem [{\citenamefont {Edwards}\ and\ \citenamefont
  {Lenard}(1962)}]{lenard2}%
  \BibitemOpen
  \bibfield  {author} {\bibinfo {author} {\bibfnamefont {S.~F.}\ \bibnamefont
  {Edwards}}\ and\ \bibinfo {author} {\bibfnamefont {A.}~\bibnamefont
  {Lenard}},\ }\bibfield  {title} {\bibinfo {title} {Exact statistical
  mechanics of a one-dimensional system with {Coulomb} forces. ii. the method
  of functional integration},\ }\href
  {https://doi.org/http://dx.doi.org/10.1063/1.1724281} {\bibfield  {journal}
  {\bibinfo  {journal} {J. Math. Phys.}\ }\textbf {\bibinfo {volume} {3}},\
  \bibinfo {pages} {778} (\bibinfo {year} {1962})}\BibitemShut {NoStop}%
\bibitem [{\citenamefont {Prager}(1962)}]{prager}%
  \BibitemOpen
  \bibfield  {author} {\bibinfo {author} {\bibfnamefont {S.}~\bibnamefont
  {Prager}},\ }\bibinfo {title} {The one-dimensional plasma},\ in\ \href
  {https://doi.org/https://doi.org/10.1002/9780470143506.ch5} {\emph {\bibinfo
  {booktitle} {Advances in Chemical Physics}}}\ (\bibinfo  {publisher} {John
  Wiley \& Sons},\ \bibinfo {year} {1962})\ pp.\ \bibinfo {pages}
  {201--224}\BibitemShut {NoStop}%
\bibitem [{\citenamefont {Baxter}(1963)}]{baxter}%
  \BibitemOpen
  \bibfield  {author} {\bibinfo {author} {\bibfnamefont {R.~J.}\ \bibnamefont
  {Baxter}},\ }\bibfield  {title} {\bibinfo {title} {Statistical mechanics of a
  one-dimensional {Coulomb} system with a uniform charge background},\ }\href
  {https://doi.org/10.1017/S0305004100003790} {\bibfield  {journal} {\bibinfo
  {journal} {Math. Proc. Cambridge Philos. Soc.}\ }\textbf {\bibinfo {volume}
  {59}},\ \bibinfo {pages} {779–787} (\bibinfo {year} {1963})}\BibitemShut
  {NoStop}%
\bibitem [{\citenamefont {Varela}\ \emph {et~al.}(2017)\citenamefont {Varela},
  \citenamefont {T\'ellez},\ and\ \citenamefont {Trizac}}]{vtt2016}%
  \BibitemOpen
  \bibfield  {author} {\bibinfo {author} {\bibfnamefont {L.}~\bibnamefont
  {Varela}}, \bibinfo {author} {\bibfnamefont {G.}~\bibnamefont {T\'ellez}},\
  and\ \bibinfo {author} {\bibfnamefont {E.}~\bibnamefont {Trizac}},\
  }\bibfield  {title} {\bibinfo {title} {Configurational and energy landscape
  in one-dimensional {Coulomb} systems},\ }\href
  {https://doi.org/https://doi.org/10.1103/PhysRevE.95.022112} {\bibfield
  {journal} {\bibinfo  {journal} {Phys. Rev. E}\ }\textbf {\bibinfo {volume}
  {95}},\ \bibinfo {pages} {022112} (\bibinfo {year} {2017})}\BibitemShut
  {NoStop}%
\bibitem [{\citenamefont {Chepelianskii}\ \emph {et~al.}(2009)\citenamefont
  {Chepelianskii}, \citenamefont {Mohammad-Rafiee}, \citenamefont {Trizac},\
  and\ \citenamefont {Rapha\"el}}]{CMTR09}%
  \BibitemOpen
  \bibfield  {author} {\bibinfo {author} {\bibfnamefont {A.~D.}\ \bibnamefont
  {Chepelianskii}}, \bibinfo {author} {\bibfnamefont {F.}~\bibnamefont
  {Mohammad-Rafiee}}, \bibinfo {author} {\bibfnamefont {E.}~\bibnamefont
  {Trizac}},\ and\ \bibinfo {author} {\bibfnamefont {E.}~\bibnamefont
  {Rapha\"el}},\ }\bibfield  {title} {\bibinfo {title} {On the effective charge
  of hydrophobic polyelectrolytes},\ }\href
  {https://doi.org/https://doi.org/10.1021/jp8076276} {\bibfield  {journal}
  {\bibinfo  {journal} {J. Phys. Chem. B}\ }\textbf {\bibinfo {volume} {113}},\
  \bibinfo {pages} {3743} (\bibinfo {year} {2009})}\BibitemShut {NoStop}%
\bibitem [{\citenamefont {Chepelianskii}\ \emph {et~al.}(2011)\citenamefont
  {Chepelianskii}, \citenamefont {Closa}, \citenamefont {Rapha\"el},\ and\
  \citenamefont {Trizac}}]{CCRT11}%
  \BibitemOpen
  \bibfield  {author} {\bibinfo {author} {\bibfnamefont {A.~D.}\ \bibnamefont
  {Chepelianskii}}, \bibinfo {author} {\bibfnamefont {F.}~\bibnamefont
  {Closa}}, \bibinfo {author} {\bibfnamefont {E.}~\bibnamefont {Rapha\"el}},\
  and\ \bibinfo {author} {\bibfnamefont {E.}~\bibnamefont {Trizac}},\
  }\bibfield  {title} {\bibinfo {title} {Effective charge of hydrophobic
  polyelectrolytes},\ }\href
  {https://doi.org/https://doi.org/10.1209/0295-5075/94/68010} {\bibfield
  {journal} {\bibinfo  {journal} {Europhys. Lett.}\ }\textbf {\bibinfo {volume}
  {94}},\ \bibinfo {pages} {68010} (\bibinfo {year} {2011})}\BibitemShut
  {NoStop}%
\bibitem [{\citenamefont {Baulin}\ and\ \citenamefont {Trizac}(2012)}]{BaTr12}%
  \BibitemOpen
  \bibfield  {author} {\bibinfo {author} {\bibfnamefont {V.~A.}\ \bibnamefont
  {Baulin}}\ and\ \bibinfo {author} {\bibfnamefont {E.}~\bibnamefont
  {Trizac}},\ }\bibfield  {title} {\bibinfo {title} {Self-assembly of spherical
  interpolyelectrolyte complexes from oppositely charged polymers},\
  }\href@noop {} {\bibfield  {journal} {\bibinfo  {journal} {Soft Matter}\
  }\textbf {\bibinfo {volume} {8}},\ \bibinfo {pages} {6755} (\bibinfo {year}
  {2012})}\BibitemShut {NoStop}%
\bibitem [{\citenamefont {Trizac}\ and\ \citenamefont
  {\v{S}amaj}(2012)}]{Varenna}%
  \BibitemOpen
  \bibfield  {author} {\bibinfo {author} {\bibfnamefont {E.}~\bibnamefont
  {Trizac}}\ and\ \bibinfo {author} {\bibfnamefont {L.}~\bibnamefont
  {\v{S}amaj}},\ }in\ \href
  {https://doi.org/https://doi.org/10.3254/978-1-61499-278-3-61} {\emph
  {\bibinfo {booktitle} {Proceedings of the International School of Physics
  {Enrico} {Fermi}}}},\ Vol.\ \bibinfo {volume} {184},\ \bibinfo {editor}
  {edited by\ \bibinfo {editor} {\bibfnamefont {C.}~\bibnamefont {Bechinger}},
  \bibinfo {editor} {\bibfnamefont {F.}~\bibnamefont {Sciortino}},\ and\
  \bibinfo {editor} {\bibfnamefont {P.}~\bibnamefont {Ziherl}}}\ (\bibinfo
  {publisher} {IOS, Amsterdam},\ \bibinfo {year} {2012})\ pp.\ \bibinfo {pages}
  {61--73}\BibitemShut {NoStop}%
\bibitem [{\citenamefont {Henderson}\ \emph {et~al.}(1979)\citenamefont
  {Henderson}, \citenamefont {Blum},\ and\ \citenamefont
  {Lebowitz}}]{henderson1}%
  \BibitemOpen
  \bibfield  {author} {\bibinfo {author} {\bibfnamefont {D.}~\bibnamefont
  {Henderson}}, \bibinfo {author} {\bibfnamefont {L.}~\bibnamefont {Blum}},\
  and\ \bibinfo {author} {\bibfnamefont {J.}~\bibnamefont {Lebowitz}},\
  }\bibfield  {title} {\bibinfo {title} {An exact formula for the contact value
  of the density profile of a system of charged hard spheres near a charged
  wall},\ }\href
  {https://doi.org/https://doi.org/10.1016/S0022-0728(79)80459-3} {\bibfield
  {journal} {\bibinfo  {journal} {J. Electroanal. Chem. and Interfacial
  Electrochem.}\ }\textbf {\bibinfo {volume} {102}},\ \bibinfo {pages} {315 }
  (\bibinfo {year} {1979})}\BibitemShut {NoStop}%
\bibitem [{\citenamefont {Henderson}\ and\ \citenamefont
  {Blum}(1981)}]{henderson2}%
  \BibitemOpen
  \bibfield  {author} {\bibinfo {author} {\bibfnamefont {D.}~\bibnamefont
  {Henderson}}\ and\ \bibinfo {author} {\bibfnamefont {L.}~\bibnamefont
  {Blum}},\ }\bibfield  {title} {\bibinfo {title} {Some comments regarding the
  pressure tensor and contact theorem in a nonhomogeneous electrolyte},\ }\href
  {https://doi.org/https://doi.org/10.1063/1.442238} {\bibfield  {journal}
  {\bibinfo  {journal} {J. Chem. Phys.}\ }\textbf {\bibinfo {volume} {75}},\
  \bibinfo {pages} {2025} (\bibinfo {year} {1981})}\BibitemShut {NoStop}%
\bibitem [{poi()}]{poissoneq}%
  \BibitemOpen
  \href@noop {} {\bibinfo {title} {Throughout this document the {Poisson}
  equation is $\phi'' = -2\rho/\epsilon $, where $\rho$ is the charge
  density.}}\BibitemShut {Stop}%
\bibitem [{\citenamefont {Kandu{\v{c}}}\ \emph {et~al.}(2008)\citenamefont
  {Kandu{\v{c}}}, \citenamefont {Trulsson}, \citenamefont {Naji}, \citenamefont
  {Burak}, \citenamefont {Forsman},\ and\ \citenamefont
  {Podgornik}}]{Kanduc2008}%
  \BibitemOpen
  \bibfield  {author} {\bibinfo {author} {\bibfnamefont {M.}~\bibnamefont
  {Kandu{\v{c}}}}, \bibinfo {author} {\bibfnamefont {M.}~\bibnamefont
  {Trulsson}}, \bibinfo {author} {\bibfnamefont {A.}~\bibnamefont {Naji}},
  \bibinfo {author} {\bibfnamefont {Y.}~\bibnamefont {Burak}}, \bibinfo
  {author} {\bibfnamefont {J.}~\bibnamefont {Forsman}},\ and\ \bibinfo {author}
  {\bibfnamefont {R.}~\bibnamefont {Podgornik}},\ }\bibfield  {title} {\bibinfo
  {title} {Weak- and strong-coupling electrostatic interactions between
  asymmetrically charged planar surfaces},\ }\href
  {https://doi.org/https://doi.org/10.1103/PhysRevE.78.061105} {\bibfield
  {journal} {\bibinfo  {journal} {Phys. Rev. E}\ }\textbf {\bibinfo {volume}
  {78}},\ \bibinfo {pages} {061105} (\bibinfo {year} {2008})}\BibitemShut
  {NoStop}%
\bibitem [{\citenamefont {Wang}\ and\ \citenamefont
  {Schiavone}(2019)}]{Wang2019}%
  \BibitemOpen
  \bibfield  {author} {\bibinfo {author} {\bibfnamefont {X.}~\bibnamefont
  {Wang}}\ and\ \bibinfo {author} {\bibfnamefont {P.}~\bibnamefont
  {Schiavone}},\ }\bibfield  {title} {\bibinfo {title} {Three-dimensional
  electric potential induced by a point singularity in a multilayered
  dielectric medium},\ }\href {https://doi.org/10.1007/s10483-019-2519-9}
  {\bibfield  {journal} {\bibinfo  {journal} {Appl. Math. and Mech.}\ }\textbf
  {\bibinfo {volume} {40}},\ \bibinfo {pages} {1327} (\bibinfo {year}
  {2019})}\BibitemShut {NoStop}%
\bibitem [{\citenamefont {Jackson}(1975)}]{jackson}%
  \BibitemOpen
  \bibfield  {author} {\bibinfo {author} {\bibfnamefont {J.~D.}\ \bibnamefont
  {Jackson}},\ }\href {https://cds.cern.ch/record/100964} {\emph {\bibinfo
  {title} {{Classical Electrodynamics; 2nd ed.}}}}\ (\bibinfo  {publisher}
  {Wiley},\ \bibinfo {address} {New York},\ \bibinfo {year} {1975})\BibitemShut
  {NoStop}%
\bibitem [{\citenamefont {Carnie}\ and\ \citenamefont
  {Chan}(1981)}]{carnie_chan}%
  \BibitemOpen
  \bibfield  {author} {\bibinfo {author} {\bibfnamefont {S.~L.}\ \bibnamefont
  {Carnie}}\ and\ \bibinfo {author} {\bibfnamefont {D.~Y.~C.}\ \bibnamefont
  {Chan}},\ }\bibfield  {title} {\bibinfo {title} {The statistical mechanics of
  the electrical double layer: Stress tensor and contact conditions},\ }\href
  {https://doi.org/https://doi.org/10.1063/1.441189} {\bibfield  {journal}
  {\bibinfo  {journal} {J. Chem. Phys.}\ }\textbf {\bibinfo {volume} {74}},\
  \bibinfo {pages} {1293} (\bibinfo {year} {1981})}\BibitemShut {NoStop}%
\end{thebibliography}%
\end{document}